\documentclass[a4paper,twocolumn, english,aps,superscriptaddress,showpacs]{revtex4}
\usepackage{times}
\usepackage[T1]{fontenc}
\usepackage[latin1]{inputenc}
\usepackage{amsmath}
\usepackage{graphicx, color}
\usepackage{amssymb}

\makeatletter

\def\ket#1{\mathinner{|{#1}\rangle}}

\usepackage{babel}
\makeatother
\begin{document}

\title{Inter-species tunneling in One-dimensional Bose mixtures}

\author{Anika C. Pflanzer}

\email{anika.pflanzer@mpq.mpg.de}

\affiliation{Physikalisches Institut, Universit\"{a}t Heidelberg, Philosophenweg
12, 69120 Heidelberg, Germany}

\author{Sascha Z\"ollner}

\email{zoellner@nbi.dk}

\affiliation{Niels Bohr International Academy, Niels Bohr Institute, Blegdamsvej
17, 2100 K\o benhavn \O, Denmark}

\author{Peter Schmelcher}

\email{peter.schmelcher@pci.uni-heidelberg.de}

\affiliation{Theoretische Chemie, Universit\"{a}t Heidelberg, INF 229, 69120
Heidelberg, Germany}

\affiliation{Physikalisches Institut, Universit\"{a}t Heidelberg, Philosophenweg
12, 69120 Heidelberg, Germany}

\begin{abstract}
We study the ground-state properties and quantum dynamics of few-boson mixtures with strong inter-species repulsion
in one-dimensional traps. If one species localizes at the center, e.g., due to a very large mass compared to the other component, it represents an effective barrier for the latter and the system can be mapped onto identical bosons in a double well. For weaker localization, the barrier atoms begin to respond to the light component, leading to an induced attraction between the mobile atoms that may even outweigh their bare intra-species repulsion.
To explain the resulting effects, we derive an effective Hubbard model for the lighter species accounting for the backaction of the barrier in correction terms to the lattice parameters. Also the tunneling is drastically affected: Varying the degree of localization of the ``barrier'' atoms, the dynamics of intrinsically noninteracting bosons can change from Rabi oscillations to effective pair tunneling. For identical fermions (or fermionized bosons) this leads to the tunneling of attractively bound pairs.
\end{abstract}

\date{\today}


\maketitle

\section{Introduction}
Ultracold atoms have become an important tool in the investigation of strongly correlated quantum systems \cite{bloch07}. Besides controlling the external degrees of freedom via optical and magnetic traps,  one can vary the atoms' interaction strengths by the use of Feshbach resonances \cite{koehler06}. In one dimension, confinement-induced resonances \cite{Olshanii1998a} enable one to uncover the whole range of interactions from strong attraction to hard-core repulsion. With this flexible toolbox at hand, it has become possible to study both the steady state and the quantum dynamics of atoms at an unpreceded level of accuracy. 
For example, optical lattices have proven to be powerful quantum simulators, giving insight into well-known condensed matter phenomena, like, e.g., the superfluid to Mott-insulator transition \cite{greiner2002, lewenstein07}.\ 
While many studies concentrate on the weakly-interacting regime, in particular, in one dimension, the strongly-interacting limit yields novel effects and valuable insights into properties of strongly-correlated systems. If \textit{single-component} bosons in one dimension repel each other infinitely strongly, the bosons' hard-core repulsion can be mimicked by Pauli's exclusion principle and accordingly, they can be mapped to noninteracting fermions \cite{girardeau60}. While the hard-core bosons are identical with the fermions in all local quantities, like the density or the energy spectrum, nonlocal properties such as the momentum distribution are very different. This phenomenon, known as the Tonks-Girardeau gas, has attracted a lot of attention \cite{bloch07}, especially after its experimental realization \cite{kinoshita04,paredes04}.\\ 
More recently, one-dimensional bosonic mixtures -- consisting of different species
 or hyperfine components -- have received much attention \cite{cazalilla03, alon06,mishra07,roscilde07,kleine08, girardeau07,zoellner08b,hao08,hao09,tempfli09} (see also Refs.
 therein.) Their rich physical properties are rooted in the interplay between intra- and inter-species forces.
 Moreover, the trapping potential, in particular the frequencies of the
 confining potentials, may be made species dependent.\ 
 While the research focus has been on static properties in most investigations, it would be thrilling to find out more about their quantum dynamics.
More specifically, we want to study a \textit{novel kind of tunneling} \cite{pflanzer09} in this paper. Most setups assume that the atoms tunnel through a \textit{classical} barrier provided by an external potential (see \cite{smerzi95, milburn97, albiez05,winkler06,foelling07, zoellner07b, zoellner07c} and Ref. therein). Only recently, studies have put the spotlight on dynamics through quantum systems, such as atomic quantum dots \cite{recati05, fischer08} and quantum barriers \cite{maschler05,larson08} in optical cavities. 
Here, we investigate tunneling through a \textit{material} barrier realized via a quasi-one-dimensional mixture of two atomic species, where each component experiences a different degree of localization. If one species is tightly confined to the center of the trap, we show that for strong inter-species repulsion
it can indeed be understood as an effective tunnel barrier for the
other one, revealing the same properties as a system with a classical barrier. As that confinement is relaxed, the ``barrier''
atoms move due to the backaction of the other species. The dramatic effect of this additional correlation on the mobile component is studied both for the ground state and for the inter-species tunneling dynamics.

Our paper is organized as follows: Sec.~\ref{model} introduces the setup. The ground-state properties of the system are discussed in Sec.~\ref{ground_state}, first for the static case, where one species is completely frozen out at the center of the trap (Sec.~\ref{statlim}), then allowing for some flexibility of the ``barrier'' atoms (Sec.~\ref{flex}), completed by an explanation of the observed effects within effective lattice models. Section \ref{tunnel} focuses on the quantum dynamics of two identical atoms interacting with one ``barrier'' atom, rounded off by the discussion of higher particle numbers in Sec.~\ref{highernumb}. 

\section{Setup} \label{model}
We consider a mixture of two distinguishable bosonic species, which we will label $\sigma=A, B$. These may correspond to different hyperfine states of one and the same species in the case of identical masses, or to two different atomic species or isotopes in which case the two constituents could also have different masses. We study quasi-one-dimensional systems, where we assume the atoms to be tightly confined in the transverse direction, such that these degrees of freedom can be integrated out.
The Hamiltonian for an arbitrary mixture of $N=N_{\mathrm{A}}+N_{\mathrm{B}}$ bosons then reads $H=\sum_{\sigma=A,B}H_{\sigma}+H_{\mathrm{AB}}$ with the effective one-dimensional single-species and inter-species Hamiltonians 
\begin{eqnarray*}
H_{\sigma}&=&\sum_{i=1}^{N_{\sigma}}\left[\frac{1}{2m_{\sigma}}p_{\sigma,i}^{2}+U_{\sigma}(x_{\sigma,i})\right]+\sum_{i<j}g_{\sigma}\delta(x_{\sigma,i}-x_{\sigma,j})\\
H_{\mathrm{AB}}&=&\sum_{i,j}g_{\mathrm{AB}}\delta(x_{\mathrm{A},i}-x_{\mathrm{B},j}).
\end{eqnarray*}
The effective one-dimensional intra- and inter-species interactions (see \cite{Olshanii1998a} for more details) can be written as 
\begin{eqnarray*}
g_{1D, \mathrm{AB}}&=&\frac{2a_{0,\mathrm{AB}}}{\mu_{\mathrm{AB}}a_{\perp, \mathrm{AB}}^2}\left( 1- C\frac{a_{0, \mathrm{AB}}}{a_{\perp, \mathrm{AB}}}\right)^{-1}\\
 g_{1D, \sigma}&=&\frac{4a_{0, \sigma}}{m_{\sigma}a_{\perp, \sigma}^2}\left( 1- C\frac{a_{0, \sigma}}{a_{\perp, \sigma}}\right)^{-1},
\end{eqnarray*}
where the transverse trapping frequency $\omega_{\perp}$ is assumed to be the same for both species. The 3D scattering lengths are denoted as $a_{0, \mathrm{AB}}$ and $a_{0,\sigma}$, while $a_{\perp, \sigma}=\sqrt{2\hbar/ m_{\sigma}\omega_{\perp}}; a_{\perp, \mathrm{AB}}=\sqrt{\hbar/ \mu_{AB}\omega_{\perp}}$ describes the transverse confinement length, $\mu_{\mathrm{AB}}=m_{\mathrm{A}}m_{\mathrm{B}}/(m_{\mathrm{A}}+m_{\mathrm{B}})$ is the reduced mass and $C \approx 1.4603$ a constant. 

In the following we will focus on harmonic trapping potentials $U_{\sigma}=\frac{1}{2}m_{\sigma}\omega_{\sigma}^2x_{\sigma}^2$ and repulsive forces $g_{\sigma},g_{\mathrm{AB}}\ge0$. Rescaling to harmonic oscillator units $a_A=\sqrt{\hbar/m_A \omega_A}$ of the ``A'' component enables us to eliminate $m_A=\omega_A=\hbar=1$ by exploiting the scaling
\begin{equation*}
 H_{\sigma}(m_{\sigma}, \omega_{\sigma}, g_{\sigma}; X_{\sigma})= \hbar \omega_A H_{\sigma}\left( \frac{m_{\sigma}}{m_A}, \frac{\omega_{\sigma}}{\omega_A}, g^{\prime}_{\sigma}; X^{\prime}_{\sigma}\right) 
\end{equation*}
with $X^{\prime}_{\sigma}\equiv X_{\sigma}/a_{\mathrm{A}}$ and $g^{\prime}_{\sigma} \equiv g_{\sigma} \sqrt{m_{\mathrm{A}}/\hbar^3\omega_{\mathrm{A}}}$, $g^{\prime}_{\mathrm{AB}} \equiv g_{\mathrm{AB}} \sqrt{m_{\mathrm{A}}/\hbar^3\omega_{\mathrm{A}}}$. This leaves us with a Hamiltonian free of any dimensionful parameters, which is very convenient for numerical simulations. In terms of the mass ratio $\alpha=\frac{m_{\mathrm{A}}}{m_{\mathrm{B}}}$ and the frequency ratio $\beta=\frac{\omega_{\mathrm{A}}}{\omega_{\mathrm{B}}}$, the rescaled Hamiltonian now reads
\begin{eqnarray*}
\!H\negmedspace=\! \sum_{i=1}^{N_{\mathrm{A}}} \left( -\frac{1}{2}\partial_{\sigma, i}^{ 2} +\frac{1}{2}x_{\sigma, i}\right)  +\sum_{i=1}^{N_B} \left( -\frac{\alpha}{2} \partial_{\mathrm{B}i}^{2}+\frac{x^{2}_{\mathrm{B}i}}{2\alpha\beta^2}\right) &\\
\!+ g_A\sum_{i<j; \sigma=\mathrm{A,B}}\delta\left( x_{\sigma i}-x_{\sigma j}\right) 
 +g_{\mathrm{AB}}\sum_{i, j}\delta\left( x_{\mathrm{B},i}-x_{\mathrm{A},j}\right),
\label{hamil}
\end{eqnarray*}
where all primes have been left out for convenience. For simplicity, let us assume that both species experience the same trapping potential ($\beta=1$) while their masses may vary. This restriction is not crucial, and experimentally large differences in the trapping potential ($\beta\approx0$) might be more feasible than large mass differences ($\alpha \approx 0$).


\section{Ground-state properties}\label{ground_state}
In order to investigate this two-component Bose mixture, we load both species into a harmonic trap of a frequency $\omega$. If the masses of both constituents are equal, $m_{\mathrm{A}}=m_{\mathrm{B}}$, the distribution of the two species throughout the trap depends on the atom number of each constituent $N_{\mathrm{\sigma}}$ and the intra- and inter-species interactions $g_{\sigma}, g_{\mathrm{AB}}$. In addition, a variable degree of localization for each species can be realized by changing the mass ratio $\alpha$. A large mass difference $m_{\mathrm{A}}\ll m_{\mathrm{B}}$  (as, e.g., for Li-Cs) leads to different confinement lengths for both species such that $a_{\mathrm{B}} \ll 1$. This results in a separation of length scales between the lighter $A$ and the heavier $B$ bosons, such that each sort only experiences an effective potential aroused by the other one. In particular, the case of strong inter-species repulsion, $g_{\mathrm{AB}}\gg 1$, is interesting, and we will concentrate on this scenario throughout this paper. In that case, the $B$ bosons act as an effective barrier for the lighter $A$ component. While this barrier remains localized at the center of the trap for very unequal masses ($\alpha \approx 0$), the $B$ bosons begin to move due to the backaction of the lighter species for larger ratios $\alpha>0$. We will show how the density profiles change for a varying confinement length starting from equal masses, $\alpha=1$, in Sec.~\ref{sec:gen.prop}. Thereafter, we will describe the limit of a static barrier, $\alpha \rightarrow 0$, in Sec.~\ref{statlim} and carry on with an investigation of the effects of varying mass ratios $\alpha >0$ in Sec.~\ref{flex}.

\subsection{Confinement-induced Demixing}\label{sec:gen.prop}
\begin{figure}
 \begin{centering}
  \includegraphics[width=0.49\columnwidth,keepaspectratio]{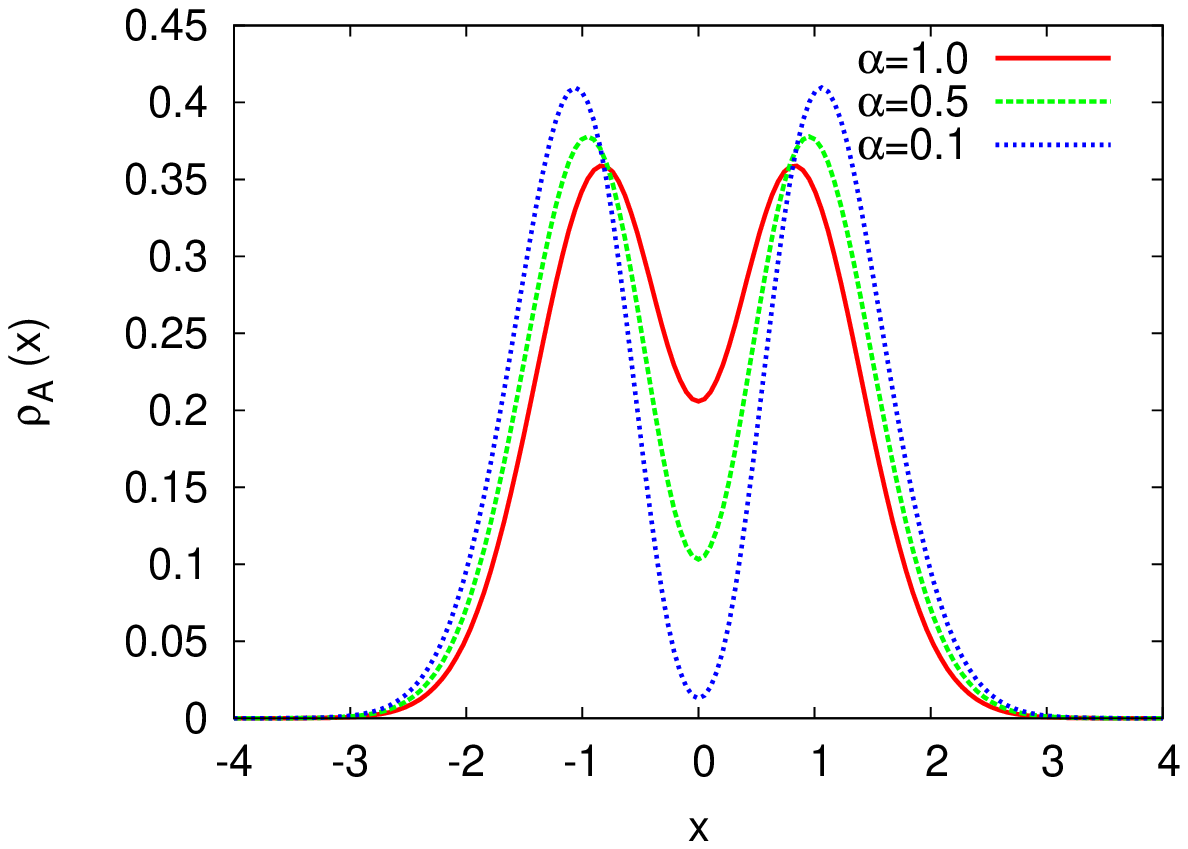}
 \includegraphics[width=0.49\columnwidth,keepaspectratio]{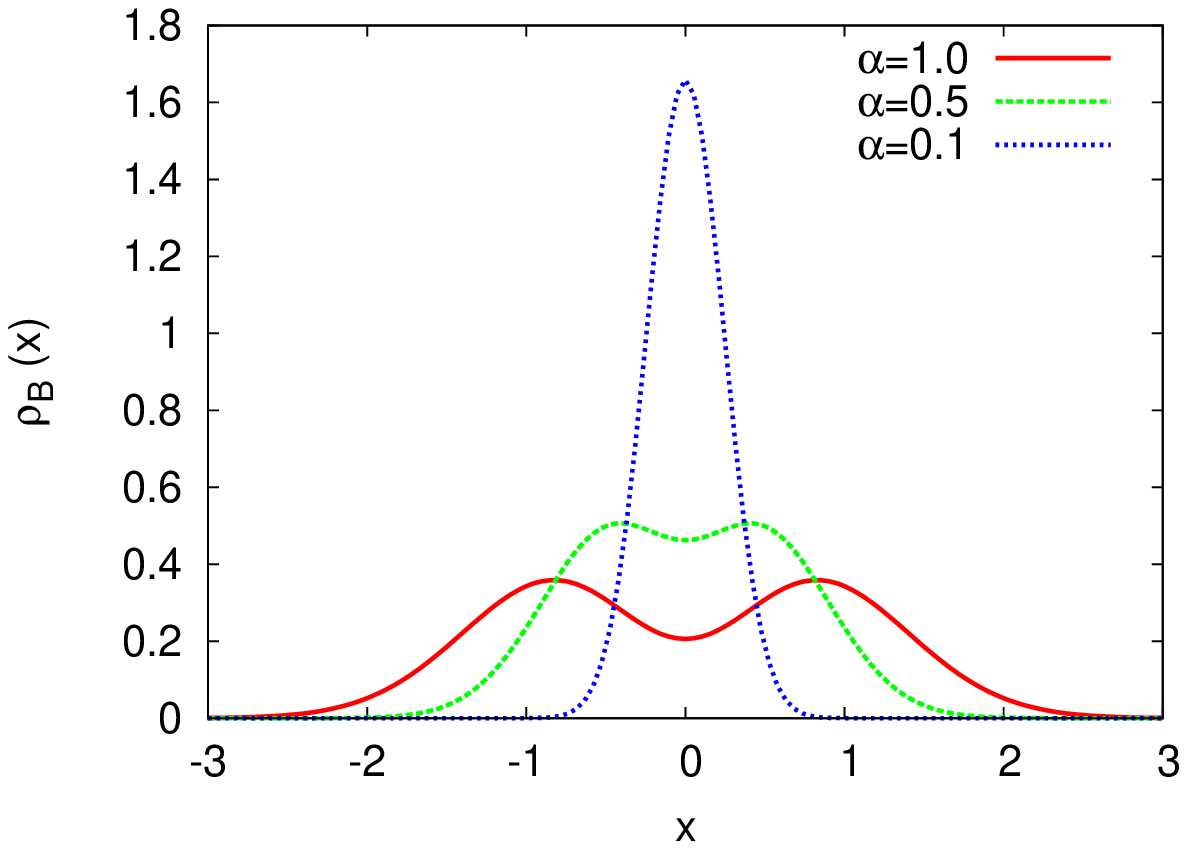}
\includegraphics[width=0.3\columnwidth, keepaspectratio]{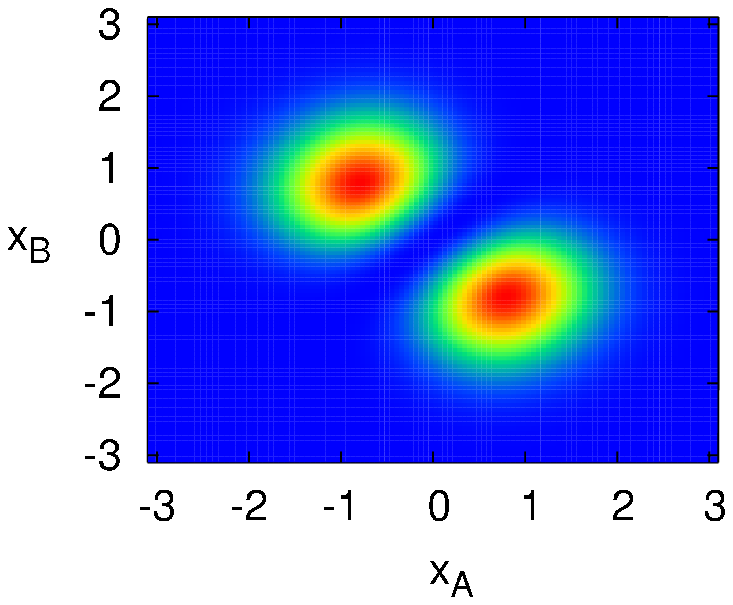}
\includegraphics[width=0.3\columnwidth, keepaspectratio]{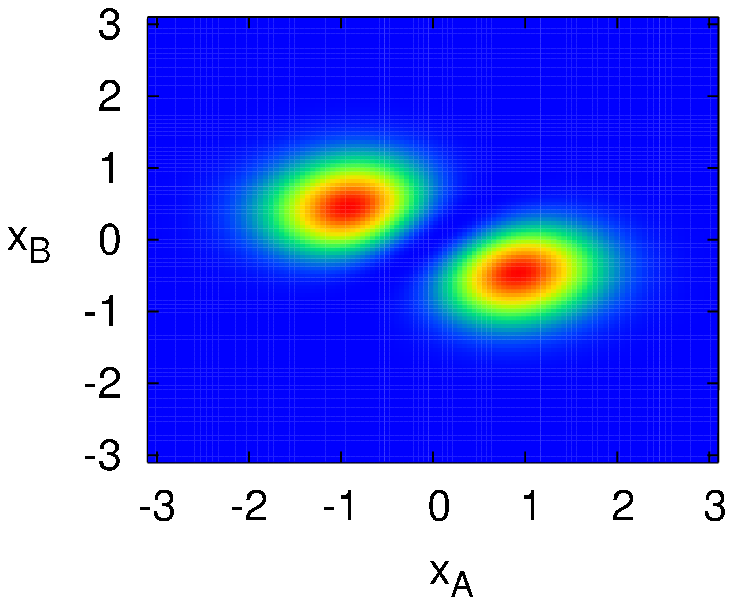}
\includegraphics[width=0.3\columnwidth, keepaspectratio]{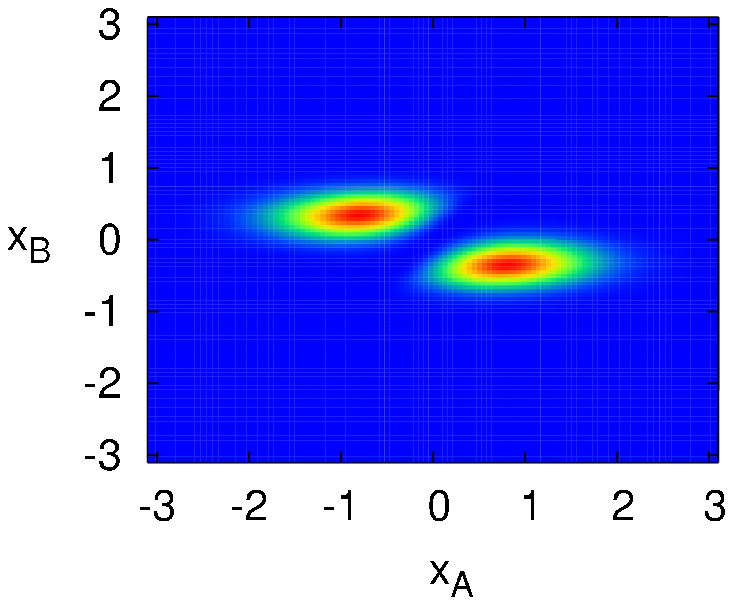}
 \end{centering}
\caption{(color online; dimensionless harmonic oscillator units $a_{\mathrm{A}}$ throughout) \textit{Top:} Influence of the mass ratio $\alpha$ on a system with  $g_{\sigma}=0$, $g_{AB}=25$, $N_A=N_B=2$. One-particle densities of the $A$ bosons (\textit{left}) and the $B$ bosons (\textit{right}) for a varying mass ratio $\alpha$ (be aware of the different scales in the plots). \textit{Bottom:} Two-body density $\rho^{(2)}_{\mathrm{AB}}(x_{1}, x_{2})$ for $\alpha=1,0.5,0.1$ (from left to right)}.
\label{Fig:densfin}
\end{figure}

Let us begin with the completely symmetric case where both species are identical in masses, external trapping and atom numbers  $\left( \alpha=\beta=1,N_{\mathrm{A}}=N_{\mathrm{B}}\right)$, which may correspond to two different internal states of the same species (e.g. two different hyperfine states). This setup is depicted for $N_A=N_B=2$ bosons in Fig.~\ref{Fig:densfin}. Accordingly, the Hamiltonian $H$ reveals complete symmetry between the bosons $A$ and $B$, resulting in identical density profiles: for both species these are spread out over the entire trap with a dip at the origin $x=0$ resulting from the strong inter-species repulsion $g_{\mathrm{AB}}=25$. This statement has to be taken with a bit of caution: the densities $\rho_{\mathrm{A}}(x)=\rho_{\mathrm{B}}(x)$ are already ensemble averages: due to the strong inter-species repulsion, in a single measurement, we will always find all atoms $N_\mathrm{A}$ at one side of the trap and all atoms $N_\mathrm{B}$ at the other side, which is depicted in the two-particle density $\rho^{(2)}_{\mathrm{AB}}(x_1, x_2)$ in Fig.~\ref{Fig:densfin}. The probability of finding an $A$ and a $B$ boson simultaneously at the same side of the trap consequently converges to zero as $g_{\mathrm{AB}} \rightarrow \infty$, and one may think of this state as a superposition of the form
$\Psi=|N_{\mathrm{A}}, 0 \rangle \otimes |0, N_{\mathrm{B}}\rangle +|0, N_{\mathrm{A}} \rangle \otimes |N_{\mathrm{B}},0 \rangle$ \cite{zoellner08b}, where $|N_{\mathrm{R}}, N_{\mathrm{L}}\rangle$ denote a state with $N_{\mathrm{R}}$ atoms at the right and $N_{\mathrm{L}}$ at the left side of the trap.

This symmetry is easily broken and the hidden phase separation becomes obvious as soon as we change the relative confinement by varying the mass ratio $\alpha$. Let us focus on the latter as depicted in Fig.~\ref{Fig:densfin}: for any $\alpha \neq 1$ the symmetry is broken and the heavier $B$ bosons experience a stronger confinement. Their oscillator length is reduced by a factor of $a_B=\sqrt{\alpha}$, preventing them from moving to the sides of the trap. For a decreasing mass ratio $\alpha$, they consequently begin to assemble more and more at the 
center of the trap, sandwiched between the lighter $A$ bosons that move to the sides. Accordingly, the dip in the density of the B bosons becomes less pronounced as $\alpha$ is decreased: while it is still visible for $\alpha=0.5$, it has vanished for $\alpha=0.1$, and the density profile of the B bosons begins to resemble a narrow Gaussian, making clear why it is appropriate to treat them as an effective barrier. On the other hand, the dip in the distribution of $A$ bosons becomes more pronounced for a decreasing $\alpha$ as a consequence of their tendency to avoid the $B$ bosons. 

\subsection{The static limit $\alpha\approx 0$}\label{statlim}
Imagine now the situation where one species (say $B$) is much more strongly localized at the center of the trap. 
As mentioned before, let us assume much heavier $B$ atoms but equal frequencies, resulting in very different length scales for the two species. Consequently, the $B$ bosons are effectively frozen at the center, while the density of the $A$ bosons changes slowly throughout the entire trap. 
To get a better understanding of this situation, let us now account for the $B$ bosons within an effective model. Due to the different masses ($\alpha \approx 0$), it is possible to carry out a Born-Oppenheimer approximation and use a product ansatz for the wave functions, such that the densities can be written as  $\hat{\rho}_{\mathrm{AB}}^{(N)}\approx\hat{\rho}_{\mathrm{A}}^{(N_{\mathrm{A}})}\otimes\hat{\rho}_{\mathrm{B}}^{(N_{\mathrm{B}})}$.
 We are now able to integrate out the $B$ component to obtain an effective Hamiltonian for the lighter species
\begin{equation}
\!\bar{H}_{A}^{(0)}\negmedspace=\! H_{\mathrm{A}}+\mathrm{tr_{B}}[H_{\mathrm{AB}}\hat{\rho}_{\mathrm{B}}^{(N_{\mathrm{B}})}]\!=\! H_{\mathrm{A}}+g_{\mathrm{AB}}\!\sum_{i}\! n_{\mathrm{B}}(x_{\mathrm{A},i})
\label{eq:H_A}.
\end{equation}
This effective single-particle potential for the $A$ bosons consists of the harmonic trapping potential and a sharp barrier at $x=0$ of width $a_{\mathrm{B}}$ given by the one-body density $\rho_{\mathrm{B}}(x)=n_{\mathrm{B}}(x)/N_{\mathrm{B}}$. Its height is proportional to the inter-species coupling and for strong repulsion $g_{\mathrm{AB}} \gg 1 $, the system bears strong resemblance to an external double well potential. We will now proceed and compare this analytical model to our numerical results in the following subsections.

\subsubsection{Density profiles}
Let us first consider the one-body densities $\rho(x)$, which describe the probability of finding one particle at position $x$. The mass ratio has been chosen as $\alpha=0.001$, giving the density of the heavier $B$ boson a $\delta$-like shape: they simply experience a harmonic oscillator potential, with the ground state being a normalized Gaussian of width $\sqrt{\alpha/2}$ (with $\hbar=1$ and $\omega=1$):
\begin{equation}
\rho_B(x_B)=\frac{1}{\sqrt{\pi \alpha}}e^{-\frac{x_{\mathrm{B}}^2}{\alpha}}
\label{eq:densb}
\end{equation}
converging towards $\delta(x_{\mathrm{B}})$ for $\alpha \rightarrow 0$. The inter-species interaction strength has been fixed at $g_{\mathrm{AB}}=25$, while the intra-species interaction $g_{\mathrm{A}}$ is varied. This barely affects the heavy $B$ bosons, which remain unchanged to a good approximation on variations of $g_{\mathrm{A}}$.\\
\begin{figure}
\begin{center}
 \includegraphics[width=0.65\columnwidth,keepaspectratio]{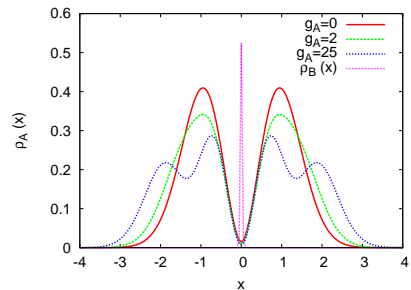}
\caption{(color online) In the limit of a vanishing mass ratio $\alpha=0.001$, the lighter $N_{\mathrm{A}}=4$ bosons develop from an uncorrelated profile (\textbf{\textcolor{red}{---}}), over a widened distribution (\textbf{\textcolor{green}{- - -}}) to a fermionized state (\textbf{\textcolor{blue}{$\cdots$}}) for an increasing intra-species interaction $g_{\mathrm{A}}$ and strong inter-species repulsion $g_{\mathrm{AB}}=25$ , while the density of the $B$ bosons represents an effective barrier for the lighter species (\textbf{\textcolor{magenta}{- - -}}) that barely changes for different $g_{\mathrm{A}}$.}
\label{dens}
\end{center}
\end{figure}

On the other hand, the strong inter-species repulsion renders the central region energetically forbidden for the lighter $A$ bosons. From their perspective, the harmonic trap is consequently split into two halves and well described by the effective model (eq.~\eqref{eq:H_A}), as we will witness in the following. 
In this sense, for $g_{\mathrm{A}}=0$ all bosons reside in the single-particle ground state $\Psi_{\mathrm{A}}=\phi_0^{\otimes N_{\mathrm{A}}}$  of the effective model, and the density distribution is consequently delocalized over the entire trap as illustrated in Fig.~\ref{dens}. If we increase $g_{\mathrm{A}}$ to $g_{\mathrm{A}} \approx 2$, the distribution is broadened due to the intra-species repulsion. For even higher interaction strengths, a completely different picture is observed: a structure of $N_{\mathrm{A}}/2$ peaks on both sides of the effective double well emerges. Physically, this means that if we measure the position of an $A$ boson, it is very likely to find it at $N_{\mathrm{A}}$ discrete spots -- and less likely to detect it anywhere in between. This \textit{localization} has a plausible physical explanation -- due to their strong repulsion, the bosons try to isolate from each other. In free space, they would simply start to move apart, until they don't feel their repulsion anymore. In our setup they cannot do so, as they are confined in a harmonic trap with an effective central barrier consisting of bosons of a different species. Therefore they try to minimize their mean interaction energy and are consequently pinpointed at more or less discrete positions. In comparison, the same phenomenon would arise for $N_A$ noninteracting fermions: due to the Pauli principle they are not allowed to occupy the same state and begin to localize in a configuration with a maximized mean distance. 
This behavior relies on the one-to-one mapping between hard-core bosons and noninteracting fermions in one-dimensional systems that has been rigorously proven in Girardeau's seminal paper in 1960 \cite{girardeau60} and is commonly referred to as \textit{fermionization} (see, e.g., \cite{zoellner06b} for more details). It relies on the fact that bosons with infinitely strong repulsion become impenetrable, which means mathematically that configuration space becomes disconnected into regions $\left\lbrace  x_i \neq x_j | i<j \right\rbrace  $, allowing us to apply the Bose-Fermi map $\Psi_{+}=A \Psi_{-}$ with $A=\prod_{i<j}\mathrm{sgn}(x_i-x_j)$. It is both valid for stationary and explicitly time-dependent states and shows that hard-core bosons and noninteracting fermions will agree in their spatial densities and energies, while nonlocal properties like the momentum distribution are different.

\subsubsection{Two-body correlations}
 \begin{figure}
\begin{center}
 \includegraphics[width=0.3\columnwidth,keepaspectratio]{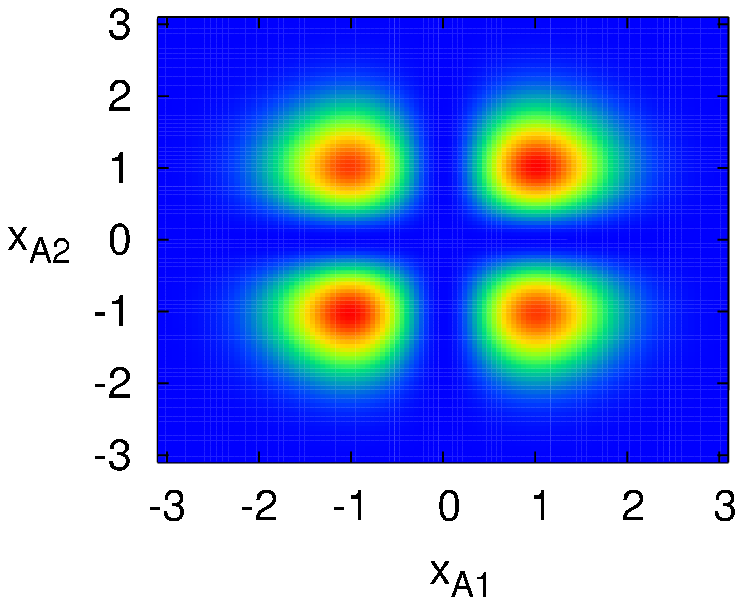}
\includegraphics[width=0.3\columnwidth,keepaspectratio]{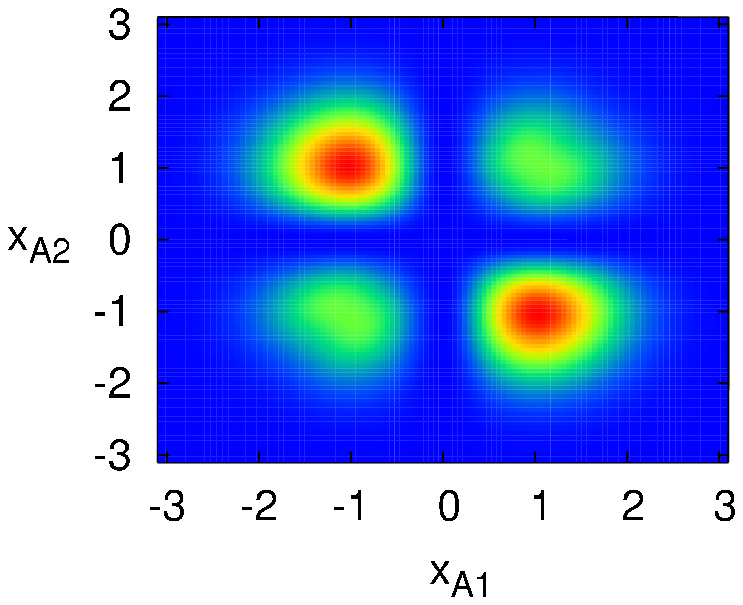}
\includegraphics[width=0.3\columnwidth,keepaspectratio]{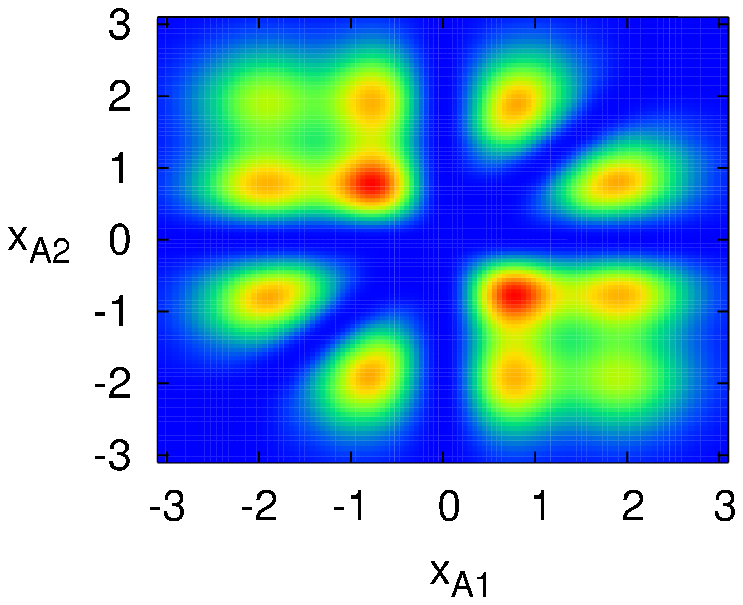}
\caption{Two-body density $\rho^{(2)}_{\mathrm{AA}}(x_{1}, x_{2})$ of the lighter $A$ species for: $N_{\mathrm{A}}=4$, $N_{\mathrm{B}}=1$ bosons, a mass ratio $\alpha=0.001$ with fixed $g_{\mathrm{AB}}=25$ and increasing $g_{\mathrm{A}}=0; 0.5; 25$ (from left to right).}
\label{corr}
\end{center}
\end{figure}
To get better insight into the mechanism underlying the fermionization crossover, we also examine the two-body densities. In the infinite-mass limit $\alpha \rightarrow 0$, $\rho_{\mathrm{A}}$ and $\rho_{\mathrm{B}}$ decouple, and we concentrate on the two-body correlation function $\rho^{(2)}_{\mathrm{AA}}(x_{1},x_{2})$. It describes the probability of finding one particle at position $x_{1}$ and the other one at $x_{2}$. For $g_{\mathrm{A}}=0$, correlations are absent and $\hat{\rho}^{(2)}_{\mathrm{AA}}=\hat{\rho}^{(1)}_{\mathrm{A}}\otimes \hat{\rho}^{(1)}_{\mathrm{A}}$. This is illustrated in the left plot in Fig.~\ref{corr}, where the depleted cross-like centered region results from the effective potential caused by the $B$ species. Besides, the $A$ bosons are coherently distributed over both ``wells''. This changes as we increase $g_{\mathrm{A}}$ to $g_{\mathrm{A}}=0.5$: a correlation hole at the diagonal $x_{\mathrm{A},1}=x_{\mathrm{A},2}$ begins to form. It has the intuitive explanation that two bosons avoid being in the same place, as this would be very costly in energy. This phenomenon is only visible in the two-body density, while it is smoothed out in the one-body density $\rho_{\mathrm{A}}^{(1)}(x)=\int dx^{\prime} \rho^{(2)}_{\mathrm{AA}}(x, x^{\prime})$ by the integration over the entire space.\\
 Tuning the interaction strength up to the intra-species fermionization limit $g_{\mathrm{A}}=25$, we witness the same picture as in the one-body density: the bosons are localized at discrete spots and the two-body density forms a checkerboard pattern. The interpretation is as follows: if one boson is fixed at a discrete spot $x_{\mathrm{A},1}$, the probability of finding a second boson at $x_{\mathrm{A},2}=x_{\mathrm{A},1}$ is zero, while finding it at three other discrete positions is equally likely. The same structure emerges for noninteracting fermions, again manifesting the equivalence between the boundary condition resulting from the strong repulsion of hard core bosons, $\Psi|_{x_{\mathrm{A},i}=x_{\mathrm{A},j}}=0$, and the Pauli principle for fermions.
\subsection{Beyond the static limit}\label{flex}
So far, we have assumed that the $B$ bosons are heavy enough to be completely frozen out at the center of the trap, and it has been shown that for $\alpha=m_{\mathrm{A}}/m_{\mathrm{B}}\approx 0$ this ``quantum barrier'' leads to the same ground-state properties as an external double well. For less restrictive mass ratios, the $B$ atoms are expected to move due to the backaction of the lighter $A$ atoms. They try to avoid the strongly repulsive lighter atoms and move towards the regions in the trap where $\rho_{\mathrm{A}}$ is smaller. Thinking of the system as an effective double well, this means that the well, in which the $A$ bosons reside, is now wider than the empty well. Hence, this renders configurations where all $A$ atoms reside at one side of the trap energetically more favorable than a setup, where the $A$ bosons are distributed equally over both sides. This additional correlation will have an immense effect on the ground state properties of the $A$ bosons. We want to investigate it in more detail and we will therefore first derive effective models valid for both weak and strong intra-species interactions $g_{\mathrm{A}}$ in Sec.~\ref{effmod} and compare it to our numerical findings thereafter in Sec.~\ref{results}.

\subsubsection{The effective model}\label{effmod}
\textit{The effective Hamiltonian}\\\label{effham}
The expansion of the Hamiltonian in orders of $\alpha$ is highly nontrivial as $H(\alpha)$ by itself does not suggest a straightforward power series in $\alpha$. In order to write the Hamiltonian in a form that is suitable for this expansion, we introduce center-of-mass and Jacobian coordinates relative to $x_{\mathrm{B}}$. Without loss of generality, we set $N_{\mathrm{B}}=1$ and transform to the new coordinate system $(r_1,...,r_{\mathrm{N_A}-1}, R)$
\begin{eqnarray*}
\!r_1\negmedspace=\!x_{\mathrm{A, 1}}-x_{\mathrm{B}};\:\:\:\:\:\:\:\:
\!r_{k}\negmedspace=\!x_{\mathrm{A,}k}-\frac{\sum_{i=1}^{k-1}m_{\mathrm{A},i}x_{\mathrm{A},i}+m_{\mathrm{B}}x_{\mathrm{B}}}{\sum_{i=1}^{k-1}m_{\mathrm{A},i}+m_{\mathrm{B}}};\\
\negmedspace \!R\negmedspace=\!\frac{\sum_{i=1}^{N_{\mathrm{A}}}m_{\mathrm{A}}x_{\mathrm{A},i}+m_{\mathrm{B}}x_{\mathrm{B}}}{\sum_{i=1}^{N_{\mathrm{A}}}m_{\mathrm{A,}i}+m_{\mathrm{B}}}\:\:\:\:\:\:\:\:\:\:\:\:\:\:\:\:\:\:\:\:\:\:\:\:
\label{jacobian}
\end{eqnarray*}
with the corresponding reduced and total masses
\begin{eqnarray*}
 \mu_{k}=\frac{\left( \sum_{i=1}^{k-1}m_{\mathrm{A},i}+m_{\mathrm{B}}\right)m_{k}}{\sum_{i=1}^{k}m_{\mathrm{A},i}+m_{\mathrm{B}}};&
M=\sum_{i=1}^{N_{\mathrm{A}}}m_{\mathrm{A},i}+m_{\mathrm{B}}.
\end{eqnarray*}
The relative Hamiltonian thus reads 
\begin {equation}
 H_{\mathrm{rel}}=\sum_k\left( \frac{p_{k}^2}{2\mu_{k}}+\frac{1}{2}\mu_{k} r_{k}^2\right) +g_{\mathrm{AB}}\sum_{k<l}\delta({c_{k,l}\left\lbrace r_{k}\right\rbrace}),
\label{Hrel}
\end {equation}
where the term containing the intra-species interaction has been left out for simplicity for now.
The last term in eq.~\eqref{Hrel} consists of combinations of relative coordinates of the form:
\begin{eqnarray}
 \nonumber g_{\mathrm{AB}}\sum_{k<l}\delta({c_{kl}\left\lbrace r_{k}\right\rbrace}) \approx g_{\mathrm{AB}}( \delta(r_{1})+\delta(r_{2}+\alpha r_{1})\\+\delta(r_{3}+\alpha r_{1} +\alpha r_{2})+...),
\label{eq:induced}
\end{eqnarray}
containing the perturbation parameter $\alpha$. A Taylor expansion of eq.~\eqref{eq:induced} leads to a Hamiltonian in orders of $\alpha$
\begin{equation*}
 H_{\mathrm{rel}}(\alpha)=H_{\mathrm{rel}}^{(0)} + H_{\mathrm{rel}}^{(1)} + H_{\mathrm{rel}}^{(2)} + O(\alpha^3)
\end{equation*}
with $H_{\mathrm{rel}}^{(0)}=\sum_k( \frac{p_{k}^2}{2\mu_{k}}+\frac{1}{2}\mu_{k} r_{k}^2)+g_{\mathrm{AB}}\sum_k \delta{(r_{k})}$,\\  $H_{\mathrm{rel}}^{(1)}= \sum_{k<l}g_{\mathrm{AB}} \alpha r_{k} \delta^{\prime}(r_l)$,\\ $H_{\mathrm{rel}}^{(2)}=\sum_{k<l}\frac{1}{2}g_{\mathrm{AB}}\alpha^2 r_{k}\delta^{\prime \prime}(r_{l})$, etc., where the prime denotes a derivative $\partial\delta(r_k)/\partial r_k$. 
Assuming that  $r_{k}\approx x_{\mathrm{A},k}-x_{\mathrm{B}}$ for small $\alpha$, we can trace out the $B$ bosons to obtain an effective Hamiltonian for the $A$ bosons:
\begin{eqnarray}
 \nonumber \bar{H}_{\mathrm{A}}(\alpha)&=& \bar{H}_{\mathrm{A}}^{(0)}+\frac{\alpha}{2}g_{\mathrm{AB}}\sum_{k\neq l}[(x_{\mathrm{A},k}-x_{\mathrm{A},l})n_{\mathrm{B}}^{\prime}(x_{\mathrm{A},l})\\&-& \nonumber n_{\mathrm{B}}(x_{\mathrm{A},l})]
+  \frac{\alpha^2}{4} g_{\mathrm{AB}}\sum_{k \neq l}[2 n_{\mathrm{B}}^{\prime}(x_{\mathrm{A},l})\\ &+&(x_{\mathrm{A},l} -x_{\mathrm{A},k})n_{\mathrm{B}}^{\prime \prime}(x_{\mathrm{A},l})]+ O(\alpha^3). \label{finalham}
\end{eqnarray}
The permutation symmetry between the $A$ bosons has been broken by the assumption $r_{k}\approx x_{\mathrm{A},k}-x_{\mathrm{B}}$ and has consequently been restored by hand in the above Hamiltonian.
While the first term in eq.~\eqref{finalham} recovers the initial infinite-mass approximation, the higher order terms may be understood as an additional external potential on $A$: $\delta U_{\mathrm{A}}(x)\equiv g_{\mathrm{AB}}\frac{N_{\mathrm{A}}-1}{2}\left\lbrace -\alpha \left[xn'_{\mathrm{B}}(x)+n_{\mathrm{B}}(x)\right]+\frac{\alpha^2}{2} \left[  2 n_{\mathrm{B}}^{\prime}(x)+x n_{\mathrm{B}}^{\prime \prime}(x)\right]\right\rbrace\\ +O(\alpha^3)$ plus an induced nonlocal interaction between two $A$ atoms: $\delta V_{\mathrm{A}}(x_{1},x_{2})\equiv \frac{g_{\mathrm{AB}}}{2} \left\lbrace \alpha \left[x_{1}n_{\mathrm{B}}^{\prime}(x_{2})+x_{2}n_{\mathrm{B}}^{\prime}(x_{1})\right]-\frac{\alpha^2}{2}\left[ x_1 n_{\mathrm{B}}^{\prime \prime}(x_2)+ x_2 n_{\mathrm{B}}^{\prime \prime}(x_1)\right] \right\rbrace\\ +O(\alpha^3)$, which adds to the local interaction $V_{\mathrm{A}}(x_{1},x_{2})\equiv g_{\mathrm{A}}\delta(x_{1}-x_{2})$.\\

\textit{An effective Bose-Hubbard model}\\\label{bosehub}
Given the multi-well (here: two sites) geometry experienced by the $A$ atoms, the use of simplified lattice models suggests itself. For sufficiently low intra-species interactions $g_A\leq 1$, an effective \textit{Bose-Hubbard model} proves to be useful. It relies on the lowest-band (limiting its use to small interaction strengths) and the tight-binding (limiting the model to deep wells) approximation, which translate in our setup into a relatively small intra-species interaction strength $g_{\mathrm{A}}$ and a sufficiently large inter-species interaction strength $g_{\mathrm{AB}}$. 
The eigenstates of the effective one-body Hamiltonian $\bar{h}_A=-\frac{1}{2}\partial_{\mathrm{A}}^2+\frac{1}{2} x_{\mathrm{A}}^2 +g_{\mathrm{AB}} n_{\mathrm{B}}(x_{\mathrm{A}})$, describing the completely localized case $\alpha \rightarrow 0$,
readily give access to a basis for this effective Hubbard model. These eigenstates are arranged in bands (in analogy to the Bloch functions for lattices) here denoted by $\beta$, consisting of doublets of symmetric $\phi^{(\beta)}_{0}(x)$ and antisymmetric $\phi^{(\beta)}_{1}(x)$ states. These delocalized functions are unsuitable for the evaluation of local properties and we apply a unitary transformation to obtain localized Wannier functions  $w^{(\beta)}_{s}(x)=\frac{1}{\sqrt{2}}\left(\phi_0^{(\beta)}(x)\pm\phi_{1}^{(\beta)}(x)\right)$ instead, where $s=\pm 1$ denotes the ``lattice'' site.
Let us now expand the effective Hamiltonian \eqref{finalham} $H_{\mathrm{A}}(\alpha)$ in its second-quantized form
\begin{eqnarray*}
\nonumber H_{\mathrm{A}}(\alpha)&=&\int dx \hat{\Psi}^{\dagger}(x)[\bar{h}_{\mathrm{A}}
+\delta U_{\mathrm{A}}(x)]\hat{\Psi}(x)\\\nonumber&+&\frac{1}{2} \int dx_1 dx_2 \hat{\Psi}^{\dagger}(x_1)\hat{\Psi}^{\dagger}(x_2)[g_{\mathrm{A}} \delta(x_1-x_2)\\ \nonumber &+&\delta V_{\mathrm{A}}(x_1, x_2)] \hat{\Psi}(x_2)\hat{\Psi}(x_1)
\label{ham_secquan}
\end{eqnarray*}
in terms of the Wannier functions $\hat{\Psi}(x)=\sum_s \hat{a}_s w_s^{(0)}(x)$, where $\hat{a}_s$ describes the annihilation operator at site $s$. We only take into account the lowest band as $g_{\mathrm{A}}$ is sufficiently small and finally obtain the effective Bose-Hubbard Hamiltonian
\begin{equation}
\bar{H}_{\mathrm{A}}^{(\mathrm{BH})}=-J^{(0)}\sum_{\langle s,s'\rangle}\hat{a}_{s}^{\dagger}\hat{a}_{s'}+\frac{u}{2}\sum_{s}\hat{n}_{s}(\hat{n}_{s}-1).
\label{eq:BHM}
\end{equation}
The first term is the hopping matrix element between nearest neighbors $J^{(0)}$, which is renormalized by $\delta U_{\mathrm{A}}$, yielding $-J^{(0)}=\langle w_{s}^{(0)}|\bar{h}_{A}+\delta U_{A}(x)|w_{s'}^{(0)}\rangle=-J^{(0)}_0+\delta J ^{(0)}$ (and correspondingly for the other bands $\beta$).
While the correction term $\delta J ^{(0)}$, tends to be small, the on-site renormalization due to $\delta V_A$, $u=\langle w_{s}^{(0)\otimes2}|V_{A}+\delta V_{A}|w_{s}^{(0)\otimes2}\rangle=u_{0}+\delta u$, can have a huge effect on the system: it induces correlations between different $A$ bosons and effectively leads to an additional intra-species coupling. 
To make this more clear, we carry out some analytical estimations by assuming a Gaussian density profile given by eq.~\eqref{eq:densb} for the B bosons and harmonic oscillator functions for the Wannier states with a displacement $x_s$ and a width $\gamma$. 
In this approximation, we find that the correction to the tunnel coupling $\delta J^{(0)}= -\sqrt{2} g_{\mathrm{AB}}\alpha^2\exp(-x_s^2/\gamma^2)/(5\gamma^3)+O(\alpha^3)$ is suppressed by a factor $\alpha^{2}$ and consequently small.
 Besides, all interaction terms involving different sites are several orders of magnitude smaller than the on-site interaction and will hence be neglected in the following. By contrast, the correction to the on-site interaction can have a remarkable effect on the system if it is of the same order as the $\alpha=0$ contribution. The most remarkable feature is that all $\delta u$ are smaller than zero for any choice of $\gamma$ and $x_s$ as it is visible in the contribution to first order in $\alpha$, $\delta u=-\frac{2}{\sqrt{\pi} \gamma^3} \alpha  g_{\mathrm{AB}}x_s^2 \exp(-x_s^2/(\alpha+\gamma^2)) +O(\alpha^2)$ . This demonstrates that the induced correlation can be interpreted as an effective intra-species attraction, increasing as $\alpha$ becomes larger.\\

\textit{An effective Multi-Band-Hubbard model}\label{fermhub}\\ 
The lowest-band approximation naturally breaks down as we go to higher interaction strengths. This becomes graphical in the fermionization limit $g_{\mathrm{A}}\rightarrow \infty$, where the hard-core bosons can be mapped onto noninteracting fermions. According to the Pauli principle these are prevented from occupying the same band and consequently, $N_{\mathrm{A}}$ bands have to be included for calculations with $N_{\mathrm{A}}$ hard-core bosons. Consequently, we can write down an effective Fermi-Hubbard model (FHM) for identical fermions:
\begin{eqnarray}
\nonumber \! \bar{H}_{\mathrm{A}}^{(\mathrm{FH})}&=&-\sum_{\langle s,s'\rangle,\beta}J^{(\beta)}\hat{f}_{s}^{(\beta)\dagger}\hat{f}_{s'}^{(\beta)}-\sum_{s, \beta}\mu^{(\beta)}_s \hat{n}^{(\beta)}_s\\ \nonumber \!&&+\frac{1}{2}\sum_s\sum_{\alpha, \beta, \alpha^{\prime}\beta^{\prime}} \delta u^{(\alpha \beta \alpha' \beta')} \hat{f}^{(\alpha)\dagger}_s\hat{f}^{(\beta)\dagger}_s\hat{f}^{(\beta^{\prime})}_s\hat{f}^{(\alpha^{\prime})}_s ,
\label{FHM}
\end{eqnarray}
where $N_A$ bands contribute as the fermions obey Pauli's exclusion principle and $\mu^{(\beta)}_s$ denotes the on-site energies. We have assumed that only nearest-neighbor tunneling gives a contribution and that inter-well interactions $\delta u_{ij}^{(\alpha \beta \alpha' \beta')}$ are negligible. The operators $\hat{f}_{s}^{(\beta)\dagger} \left( \hat{f}_{s}^{(\beta)}\right) $ create (annihilate) a particle in the $\beta$th band at the $s$th site, and its tunnel coupling is given by  $-J^{(\beta)}=\langle w_{s}^{(\beta)}|\bar{h}_{A}+\delta U_{A}(x)|w_{s'}^{(\beta)}\rangle$. The on-site interaction is given by $\delta u^{(\alpha\beta\alpha'\beta')}=\langle w^{(\alpha)}w^{(\beta)}| \delta V_A | w^{(\alpha')}w^{(\beta')}\rangle$ and the total induced interaction between different bands is quite elaborate containing all different combinations of creation and annihilation operators of the $N_{\mathrm{A}}$ bands. In order to make this term more comprehensible we write it in a different form
\begin{eqnarray*}
\nonumber \!H_{\mathrm{int}}&=&\frac{1}{2}\sum_{s, \alpha < \beta} \delta v^{(\alpha \beta\alpha\beta)} \hat{n}_s^{(\alpha)}\hat{n}_s^{(\beta)}\\ &+&\nonumber \frac{1}{2}\sum_{\alpha \neq \beta \neq \beta'}\delta v ^{(\alpha \beta \alpha \beta')}\hat{n}^{(\alpha)} \hat{f}_{s}^{(\beta)\dagger}\hat{f}_{s}^{(\beta')}\\ &+&\frac{1}{2}\sum_{\alpha \neq \beta \neq \alpha' \neq \beta'}\delta v^{(\alpha \beta \alpha' \beta')} \hat{f}_{s}^{(\alpha)\dagger}\hat{f}_{s}^{(\beta)\dagger}\hat{f}_{s}^{(\beta')}\hat{f}_{s}^{(\alpha')},
\end{eqnarray*}
with the induced inter-band interaction strengths 
\begin{eqnarray*} \!\delta v^{(\alpha \beta \alpha' \beta')}\negmedspace=\!\langle w^{(\alpha)}w^{(\beta)}| \delta V_A | w^{(\alpha')}w^{(\beta')}\rangle \\ \!+\langle w^{(\beta)}w^{(\alpha)}| \delta V_A | w^{(\beta')}w^{(\alpha')}\rangle-\langle w^{(\beta)}w^{(\alpha)}| \delta V_A | w^{(\alpha')}w^{(\beta')}\rangle\\ \!-\langle w^{(\alpha)}w^{(\beta)}| \delta V_A | w^{(\beta')}w^{(\alpha')}\rangle.
\end{eqnarray*}
In this form, it becomes clear that in the case of two $A$ bosons the interaction term can be reduced to
$H_{\mathrm{on-site}}=\sum_s \delta v^{(0101)} \hat{n}_s^{(0)}\hat{n}_s^{(1)}$.
In analogy to the bosonic case, we carry out some analytical estimations in a Wannier basis constructed of harmonic oscillator functions. While the corrections to $J^{(\beta)}$ are small (see previous paragraph), the induced $\delta v^{(\alpha \beta \alpha' \beta')}$ can become remarkable for sufficiently large $\alpha$. The induced interactions are again negative, which means that the fermionized bosons effectively attract each other. The first order contribution between the lowest bands is of the form $\delta v^{(0101)}=g_{\mathrm{AB}}\alpha x_s^2\exp(-x_s^2/\gamma^2)(-2x_s^2+\gamma^2)/(\sqrt{\pi}\gamma^5) +O(\alpha^2)$, which is negative for an appropriate choice of the lattice parameters, i.e. $x_s>\gamma/\sqrt{2}$. Of course, this induced attraction does not violate the Pauli principle as the fermions still reside in different energy bands.
 
\subsubsection{Results}\label{results}
\begin{figure}
 \includegraphics[width=0.3\columnwidth,keepaspectratio]{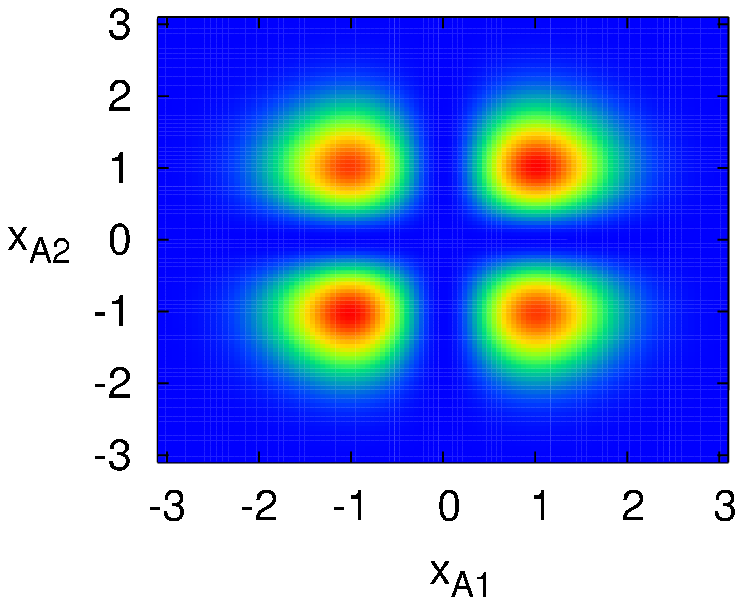}
 \includegraphics[width=0.3\columnwidth,keepaspectratio]{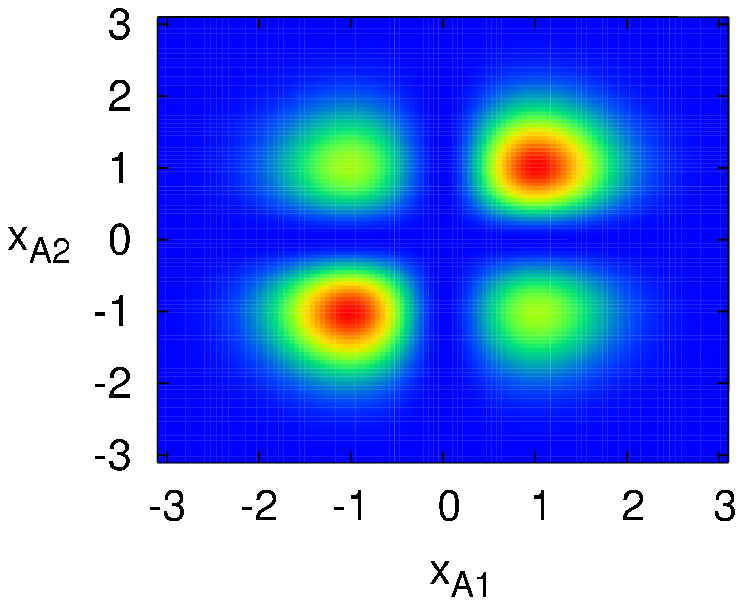}
 \includegraphics[width=0.3\columnwidth,keepaspectratio]{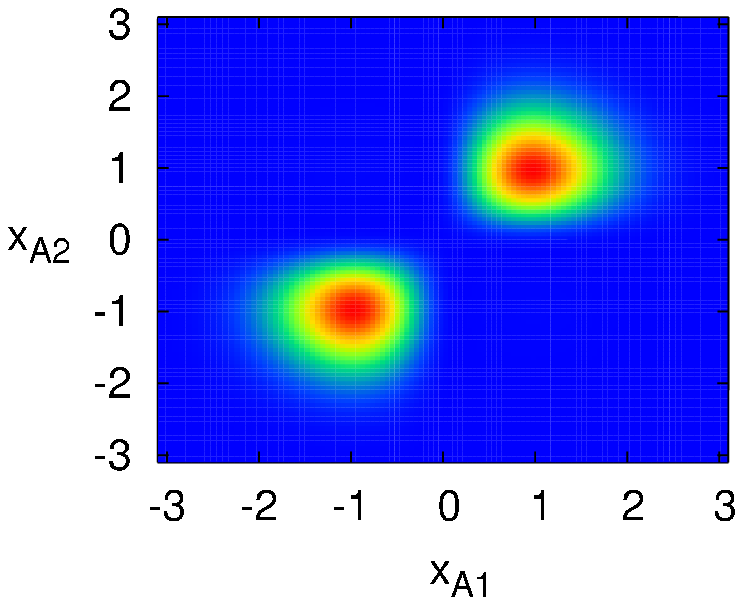}
 \includegraphics[width=0.3\columnwidth,keepaspectratio]{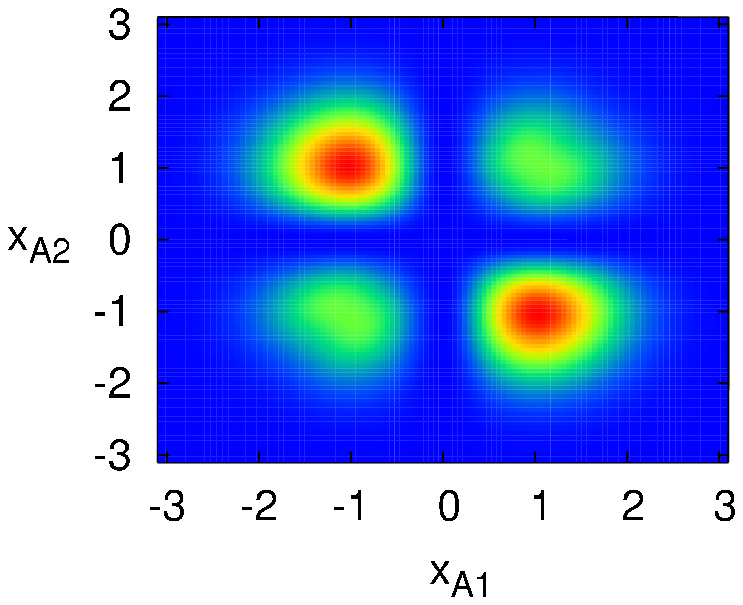}
 \includegraphics[width=0.3\columnwidth,keepaspectratio]{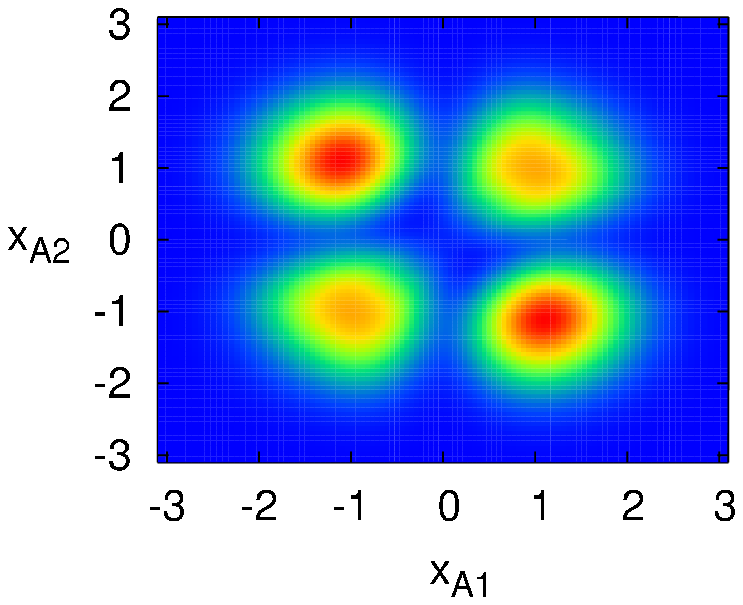}
 \includegraphics[width=0.3\columnwidth,keepaspectratio]{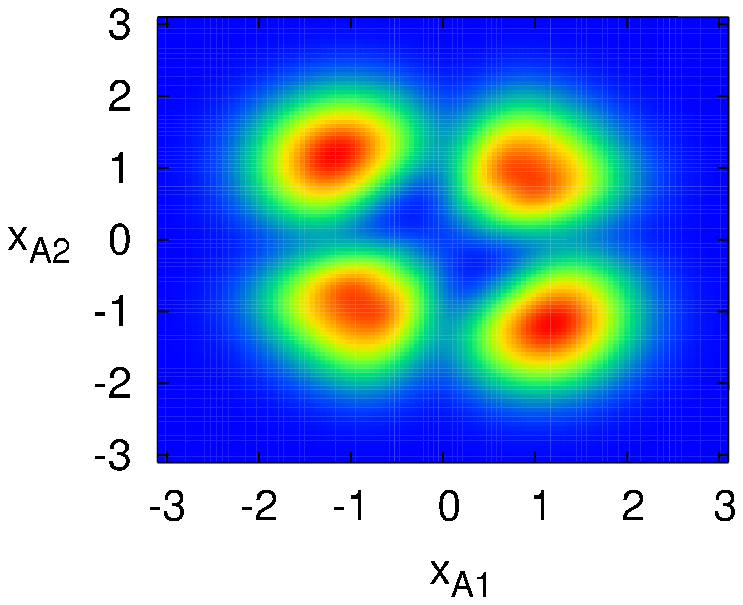}
\caption{(color online) Beyond the static ``barrier``: Two-body density $\rho_{\mathrm{AA}}^{(2)}(x_{1},x_{2})$ for $N_{\mathrm{A}}=4$ and $N_{\mathrm{B}}=1$ bosons 
for (\emph{top}) $g_{\mathrm{A}}=0$, with increasing mass ratios $\alpha=0.001,\,0.002,\,0.02$
(from left to right); (\emph{bottom}) $g_{\mathrm{A}}=0.5$, $\alpha=0.001,\,0.12,\,0.2$.}
\label{corra0}
\end{figure}

The effective Hubbard models suggest corrections to the ground state resulting from the induced intra-species interactions $\delta u$. Let us now examine how these induced effects manifest in the one- and two-body densities. While the inter-species interaction is fixed at a large repulsive value $g_{\mathrm{AB}}=25$, the intra-species interaction is varied. 
Let us first consider the limit with vanishing intra-species interactions $g_{\mathrm{A}}=0$. In particular, the two-body correlation function $\rho^{(2)}_{\mathrm{AA}}(x_{1}, x_{2})$ gives some interesting insight into these systems (see Fig.~\ref{corra0}). For $\alpha\approx 0$, there are no correlations between the $A$ bosons and the two-body density looks exactly like in the case of an external barrier. This picture changes dramatically as we vary the $\alpha \approx 0$ requirement: already for relatively small mass ratios, correlations are induced, making it more favorable for the bosons to assemble at the same site and the population of the off-diagonal decreases. Already for values as small as $\alpha=0.02$, the $A$ bosons cluster and the probability of finding them at different sites converges to zero. The effective Bose-Hubbard model in eq.~\eqref{eq:BHM} is at hand with an explanation for this phenomenon: the induced attractive interaction $\delta u $ between the $A$ bosons makes it energetically more favorable to assemble several bosons at the same site. The resulting state can be thought of as a maximally entangled or ``cat state'' where the measurement of one boson determines the outcome of all preceding measurements: $|\Psi_{\mathrm{A}}\rangle=\frac{1}{\sqrt{2}}\left( |N_{\mathrm{A}}, 0\rangle +|0, N_{\mathrm{A}}\rangle\right)$, as the entire ensemble assembles either on the left or the right side.

For larger interaction strengths $g_{\mathrm{A}}=0.5$, the system is in a ``Mott-insulating''-type state for tiny mass ratio $\alpha=0.001$. Having witnessed the effect of induced attractions for $g_{\mathrm{A}}=0$, the question naturally arises whether it might be possible to compensate the repulsion and bring the system back into an uncorrelated state. Considering the two-body correlations in Fig.~\ref{corra0} (bottom), this picture is indeed confirmed: starting from an anticorrelated ``insulating'' ground state for $\alpha\approx 0.001$, the population of the diagonal gradually increases until we end up with a seemingly uncorrelated state for $\alpha\approx 0.2$. This behavior is predicted by our effective Hubbard model: $\delta V_{\mathrm{A}}$ gives rise to an induced attraction $\delta u \approx u_0$ which can compensate or even outweigh the ``real'' intra-species interaction $V_{\mathrm{A}}$.

\begin{figure}
\includegraphics[width=0.49\columnwidth,keepaspectratio]{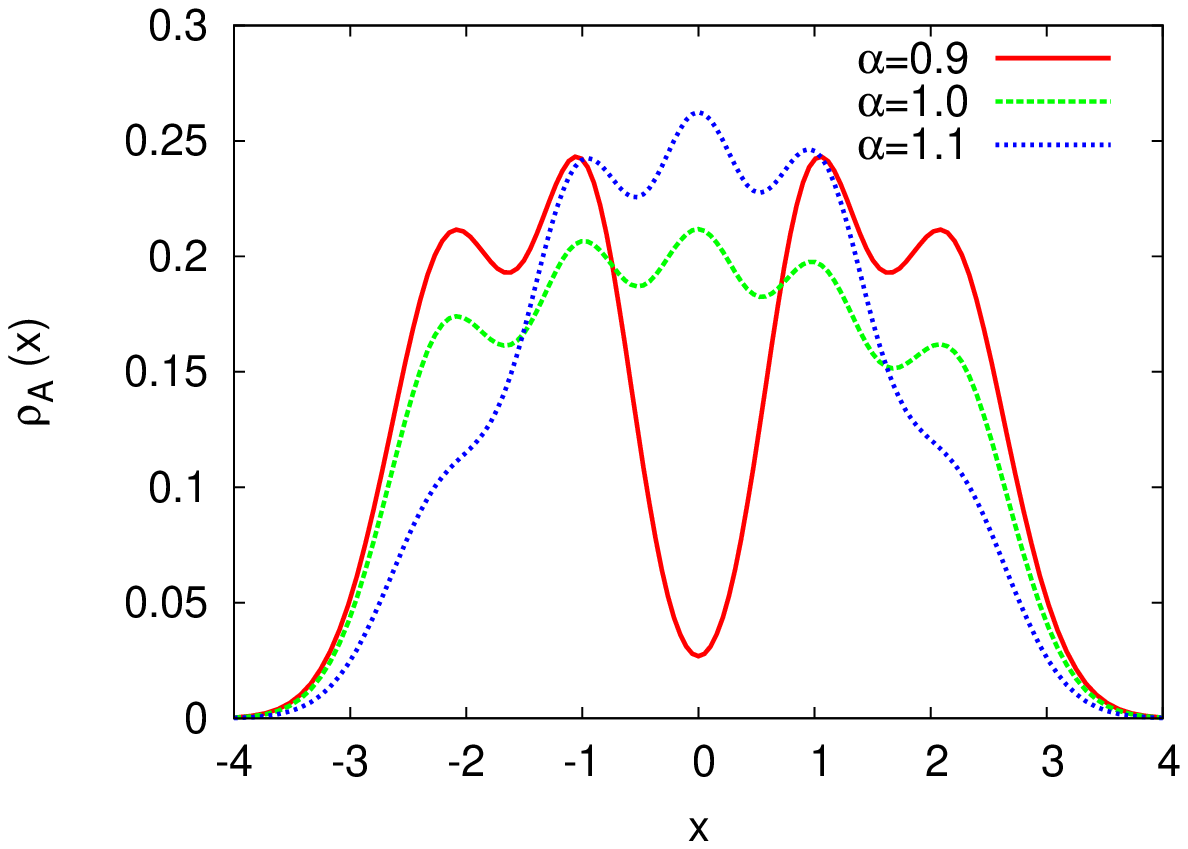}
\includegraphics[width=0.49\columnwidth,keepaspectratio]{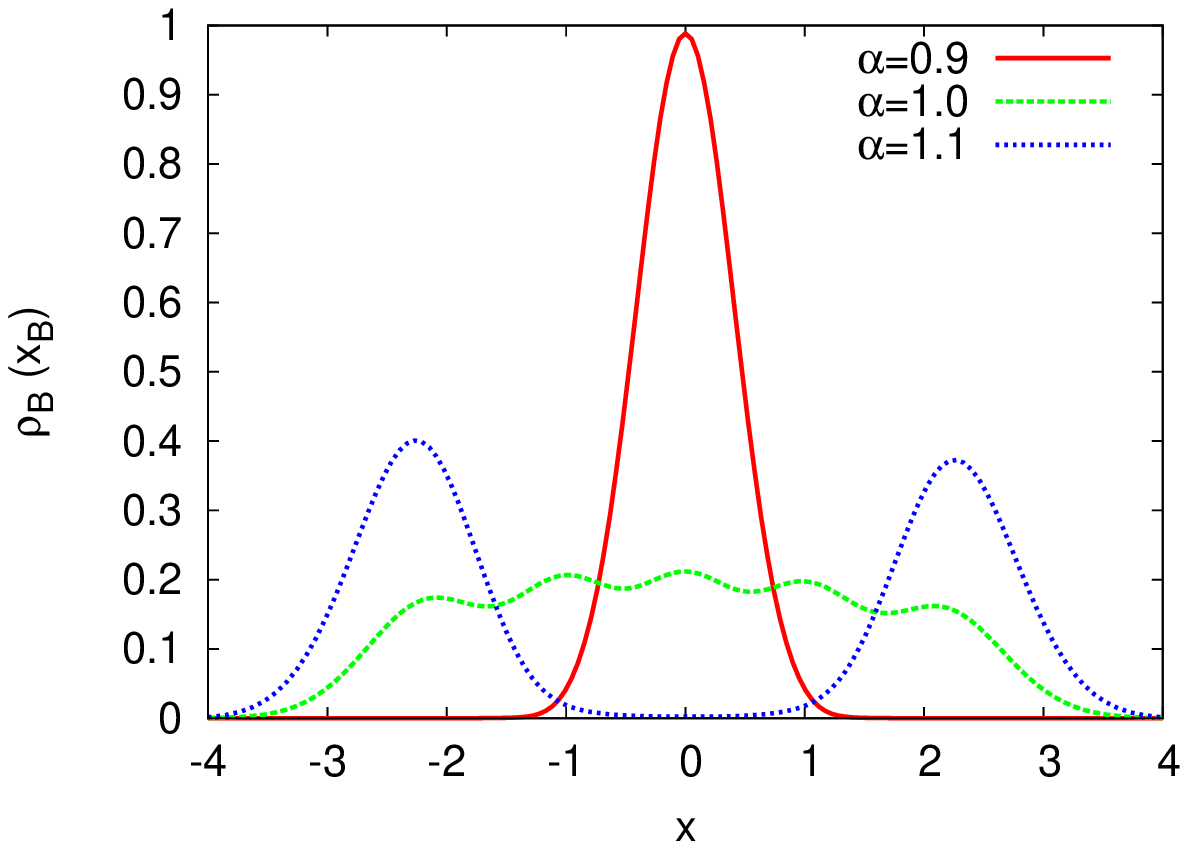}
 \includegraphics[width=0.3\columnwidth,keepaspectratio]{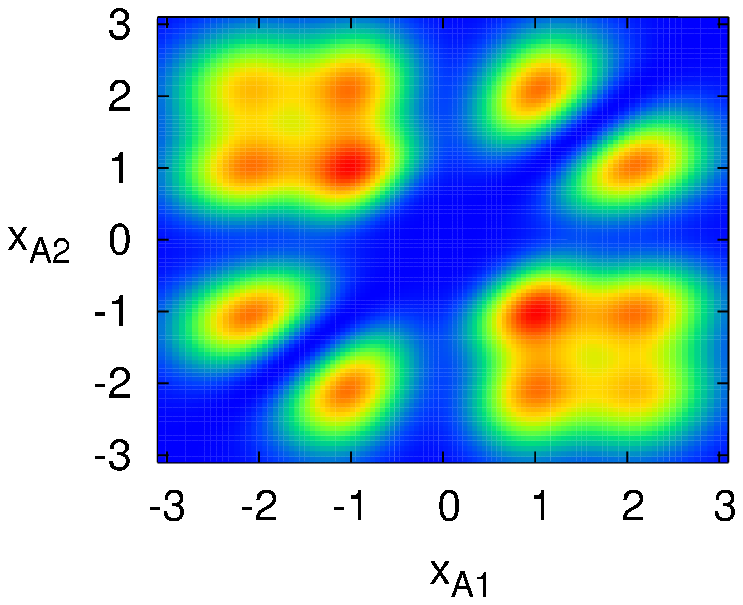}
 \includegraphics[width=0.3\columnwidth,keepaspectratio]{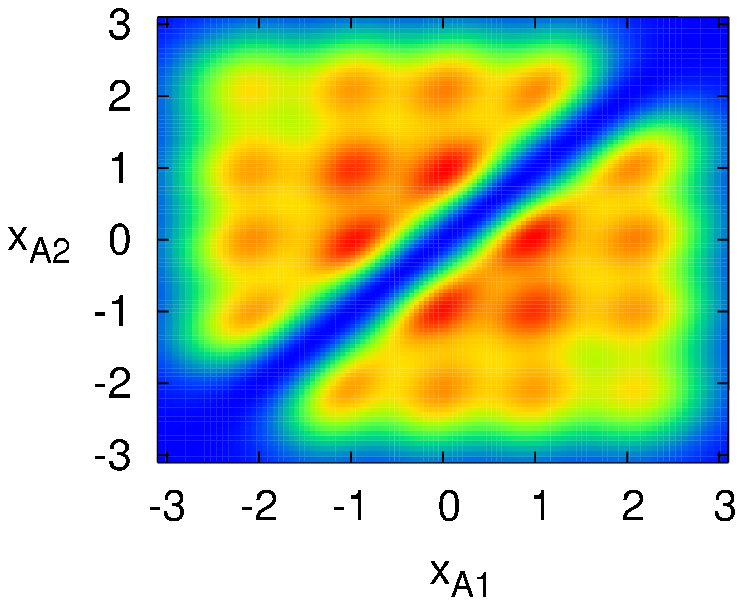}
 \includegraphics[width=0.3\columnwidth,keepaspectratio]{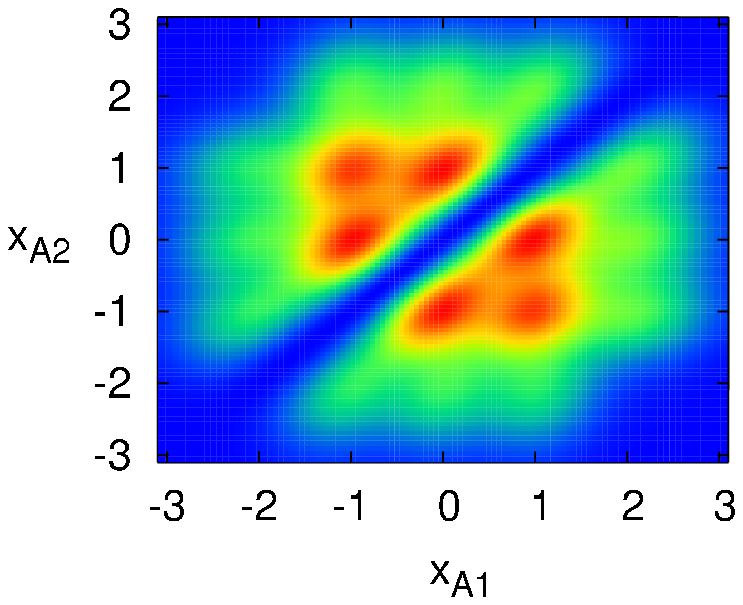}
\caption{Evolution of a system with  $g_{\mathrm{A}}=25$, $g_{\mathrm{AB}}=25$, $N_{\mathrm{A}}=4, N_{\mathrm{B}}=1$; \textit{Top}: One-particle densities for increasing $\alpha$; \textit{Left}: of the $A$ bosons, \textit{Right}: of the $B$ bosons (be aware of the different scales in the plots); \textit{Bottom}: Two-body correlation function $\rho^{(2)}_{\mathrm{AA}}(x_{1}, x_{2})$ $\alpha=0.9; 1; 1.1$ (from left to right).}
\label{fermmass}
\end{figure}
Let us now move on to the strongly repulsive fermionization limit. This case is well characterized by the effective Fermi-Hubbard model in eq.~\eqref{FHM} for small mass ratios $\alpha$. It predicts induced correlations between fermions in different bands. In order to be measurable in the ground state, this induced interaction has to be of the same order as the band gap $\epsilon^{(1)}-\epsilon^{(0)}$, making it favorable for a second fermion to occupy the same well instead of the other empty well. 
The induced interaction is naturally tiny for small $\alpha$ as illustrated in the spectrum in Fig.~\ref{spectrum_g0} and it only leads to corrections much smaller than the band gap. Therefore, no effect is visible in the ground state for relatively little mass ratios, and we will consequently look at the evolution of the states for larger $\alpha$.
 One interesting question is: What happens to a mixture of two initially fermionized components ($g_{\mathrm{A}}=25$) as their mass ratio $\alpha$ tends to one? For $\alpha=1$ with $g_{\sigma}=g_{\mathrm{AB}}=25$, the two-component mixture is related to a single-component hard-core Bose gas, that in turn can be mapped to a single-component ideal Fermi gas via the Bose-Fermi mapping \cite{girardeau60, girardeau07, zoellner08b,hao08,hao09}. The density profile is therefore the same as for an ideal single-component Fermi gas (see Fig.~\ref{fermmass}): for five hard-core bosons $N=N_{\mathrm{A}}+N_{\mathrm{B}}=5$ humps emerge. This behavior also manifests itself in the two-body correlation function: if one boson of either species $A$ or $B$ is measured at a discrete spot, the probability of finding a second boson at the same place is zero and there are four equally likely discrete positions for the remaining bosons. This special case of equal masses and interactions is highly degenerate: there is a permutation degeneracy between $A$ and $B$ bosons, and moreover in the limit $g_{\mathrm{AB}} \rightarrow \infty$, the ground state wave function also degenerates with the case, where both $A$ and $B$ are fermionic by the Bose-Fermi mapping. 
Due to the symmetry between the two species the correlation functions $\rho^{(2)}_{\mathrm{AB}}(x_1, x_2)=\rho^{(2)}_{\mathrm{AA}}(x_{1}, x_{2})$ are identical \cite{zoellner08b}.

This degeneracy is lifted as we introduce a small difference between the $A$ and $B$ bosons, for example by varying the mass ratio $\alpha$. Even for small deviations from $\alpha=1$, the symmetry in the Hamiltonian between $A$ and $B$ is broken. A slightly heavier $B$ component (see $\alpha=0.9$ in Fig.~\ref{fermmass}) already localizes at the center of the trap and  the lighter $A$ bosons move to the outer parts. This is due to the high intra-species repulsion between the $A$ bosons, which overwhelms the potential energy: even if the density of the $A$ bosons is higher ($N_{\mathrm{A}}=4$) and it is thus more expensive for them to move to the sides of the trap than it is for the $B$ boson, their strong intra-species repulsion forces them to do so, sandwiching the $B$ boson at the center. Further decreasing the mass ratio $\alpha$ has no qualitative effect on the system: the $B$ bosons remain localized at the center of the trap with a decreasing oscillator length $a_{\mathrm{B}}=\sqrt{\alpha}$, while the $A$ bosons, preserving the checkerboard pattern, tend to move slightly to the inside of the trap (not shown). 
In the opposite case $\alpha>1$, the $B$ boson, as the lighter species, moves to the sides of the trap, surrounding the $A$ bosons, that are concentrated at the center forming three wiggles (see $\alpha=1.1$ in Fig.~\ref{fermmass}). For the $A$ atoms, the presence of the heavier $B$ bosons at the sides of the trap amounts to hard-wall boundary conditions imposed via the strong repulsion between the two species. The harmonic potential has a minimum at $x=0$, and the hard-core bosons of species $A$ therefore localize around the central region and form $N_{\mathrm{A}}-1=3$ wiggles. The fourth $A$ boson is delocalized over the entire allowed region due to symmetry requirements.\\

\section{Inter-species Tunneling Dynamics}\label{tunnel}
Having investigated the equilibrium properties of a bosonic mixture where the heavier of two species acts as an effective barrier for the lighter one, it would now be interesting to learn about the nonequilibrium quantum dynamics of this setup. Do the lighter atoms tunnel through the heavier ones and how is this motion influenced by the backaction of the ``barrier'' for varying mass ratios? 
To tackle these questions, we prepare the barrier bosons $B$ at the center of the trap while the lighter $A$ atoms are entirely loaded into the left-hand side. This can be achieved by either displacing the trap center for the lighter species or by blocking the right-hand side by shining blue-detuned laser light on it. At release time $t=0$, the asymmetry in the potential of the lighter atoms is lifted and they begin to tunnel through species $B$, that react and in turn modify the effective potential of the lighter bosons. To explore the tunneling dynamics of the $A$ bosons through the barrier atoms, we fix the inter-species interaction strength at a large repulsive value $g_{\mathrm{AB}}=8$ and vary the intra-species interaction $g_{\mathrm{A}}\geq 0$, covering the full fermionization crossover.

\subsection{The noninteracting limit $g_{\mathrm{A}}=0$}
 \begin{figure}
  \includegraphics[width=0.3\columnwidth,keepaspectratio]{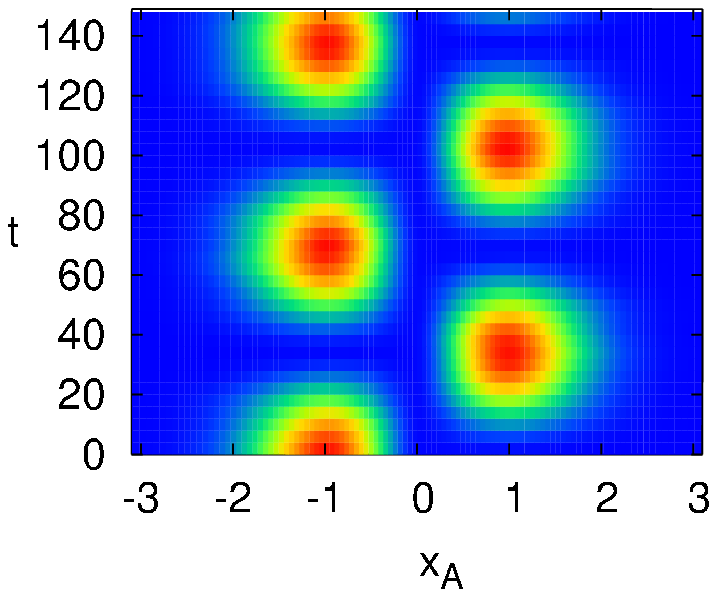}
 \includegraphics[width=0.3\columnwidth,keepaspectratio]{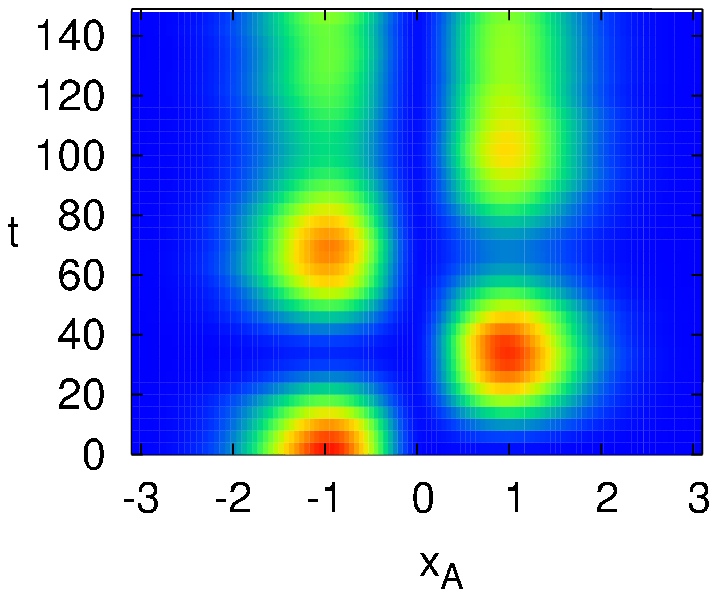}
 \includegraphics[width=0.3\columnwidth,keepaspectratio]{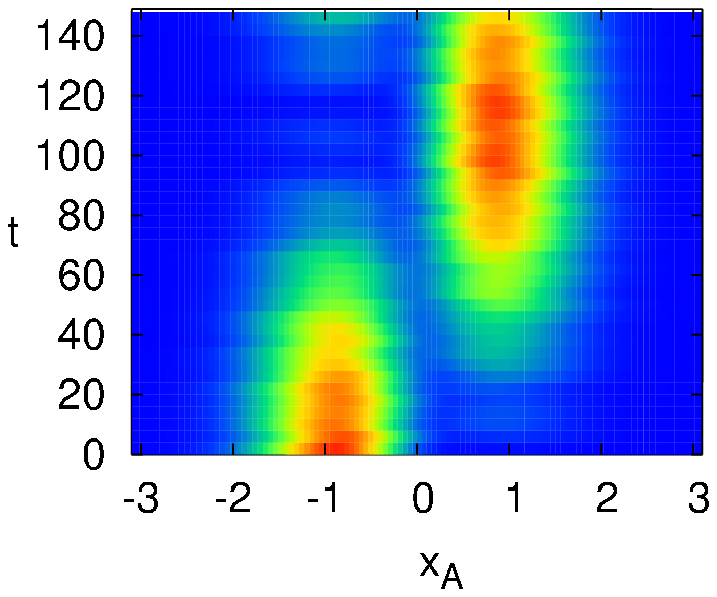}
  \includegraphics[width=0.3\columnwidth,keepaspectratio]{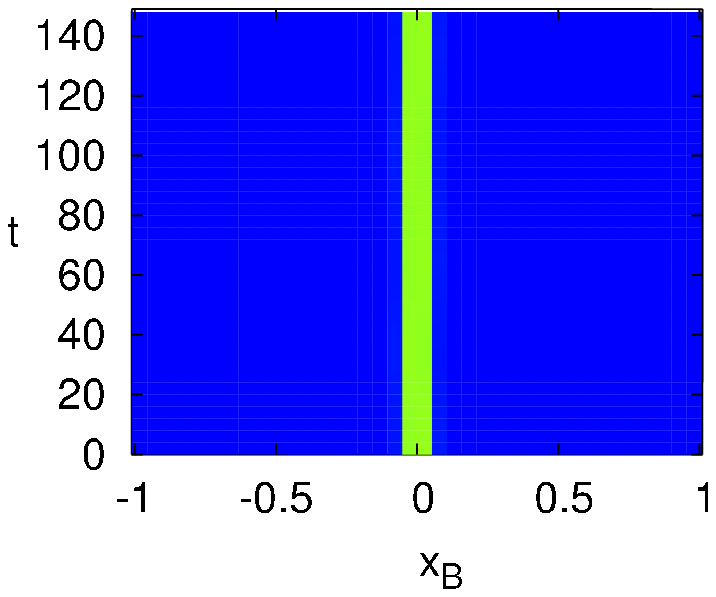}
\includegraphics[width=0.3\columnwidth,keepaspectratio]{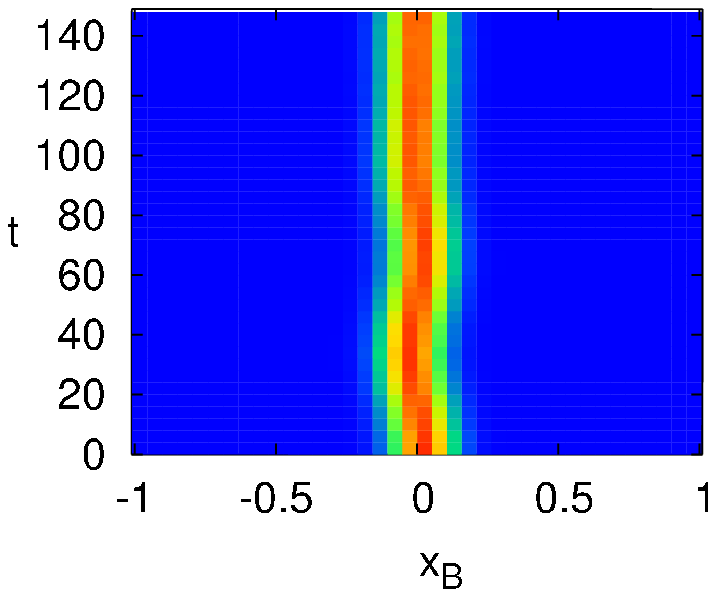}
 \includegraphics[width=0.3\columnwidth,keepaspectratio]{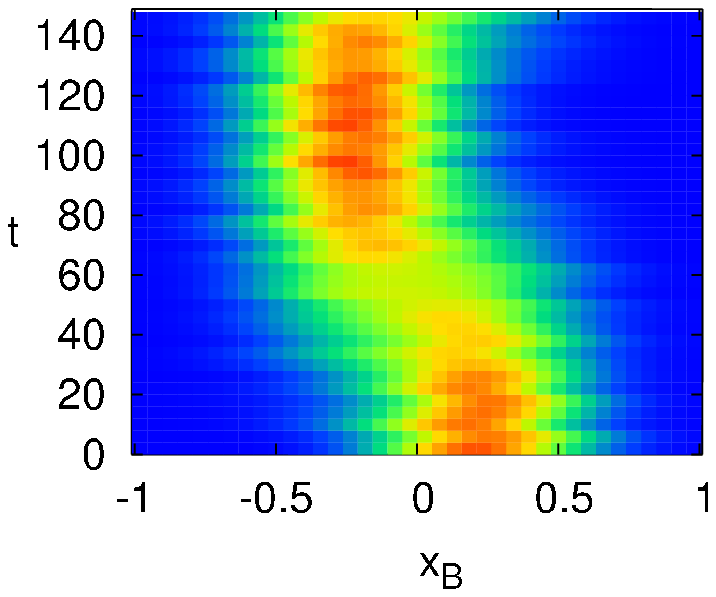}
 \includegraphics[width=0.3\columnwidth,keepaspectratio, angle=270]{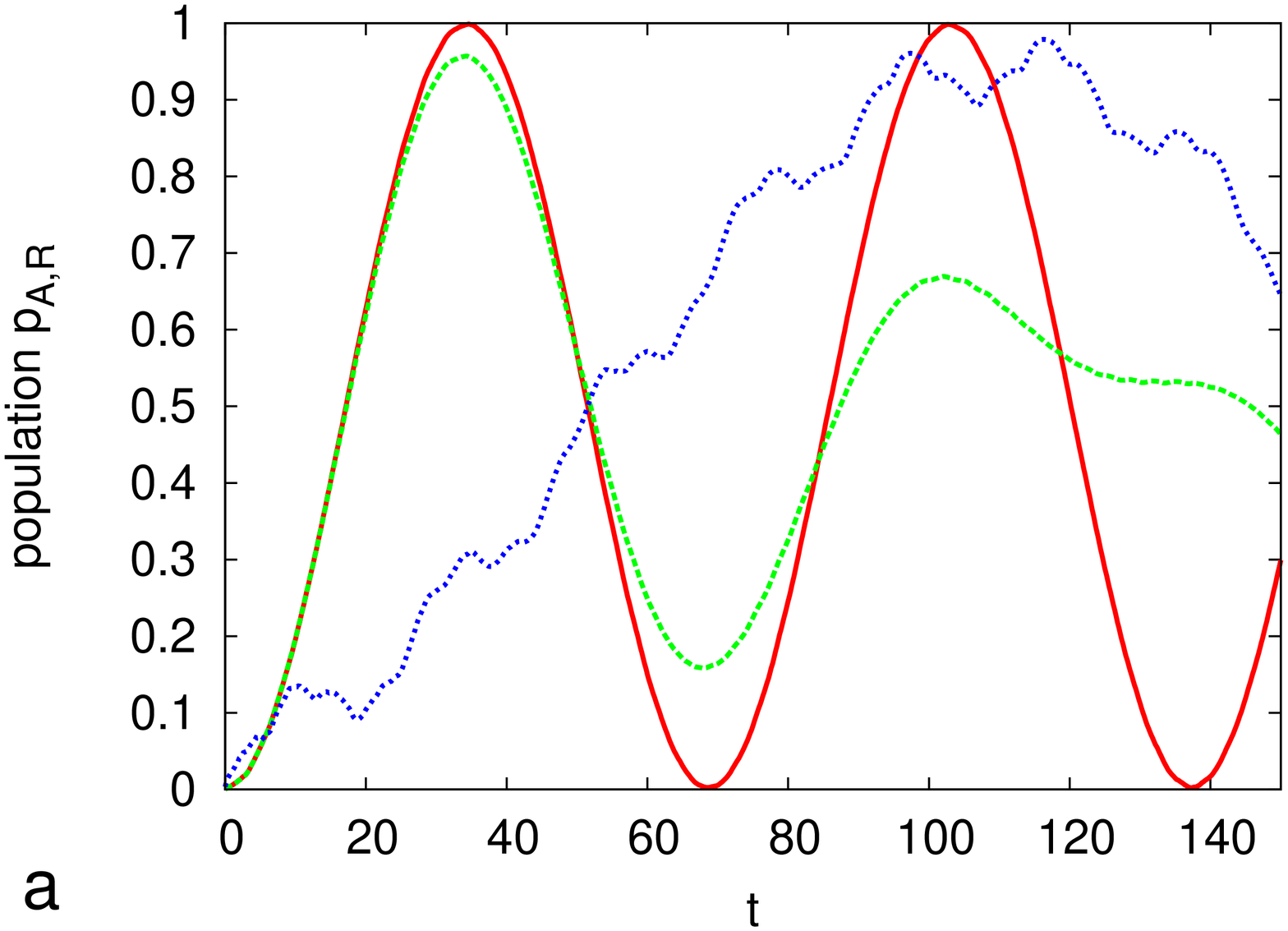}
 \includegraphics[width=0.3\columnwidth,keepaspectratio, angle=270]{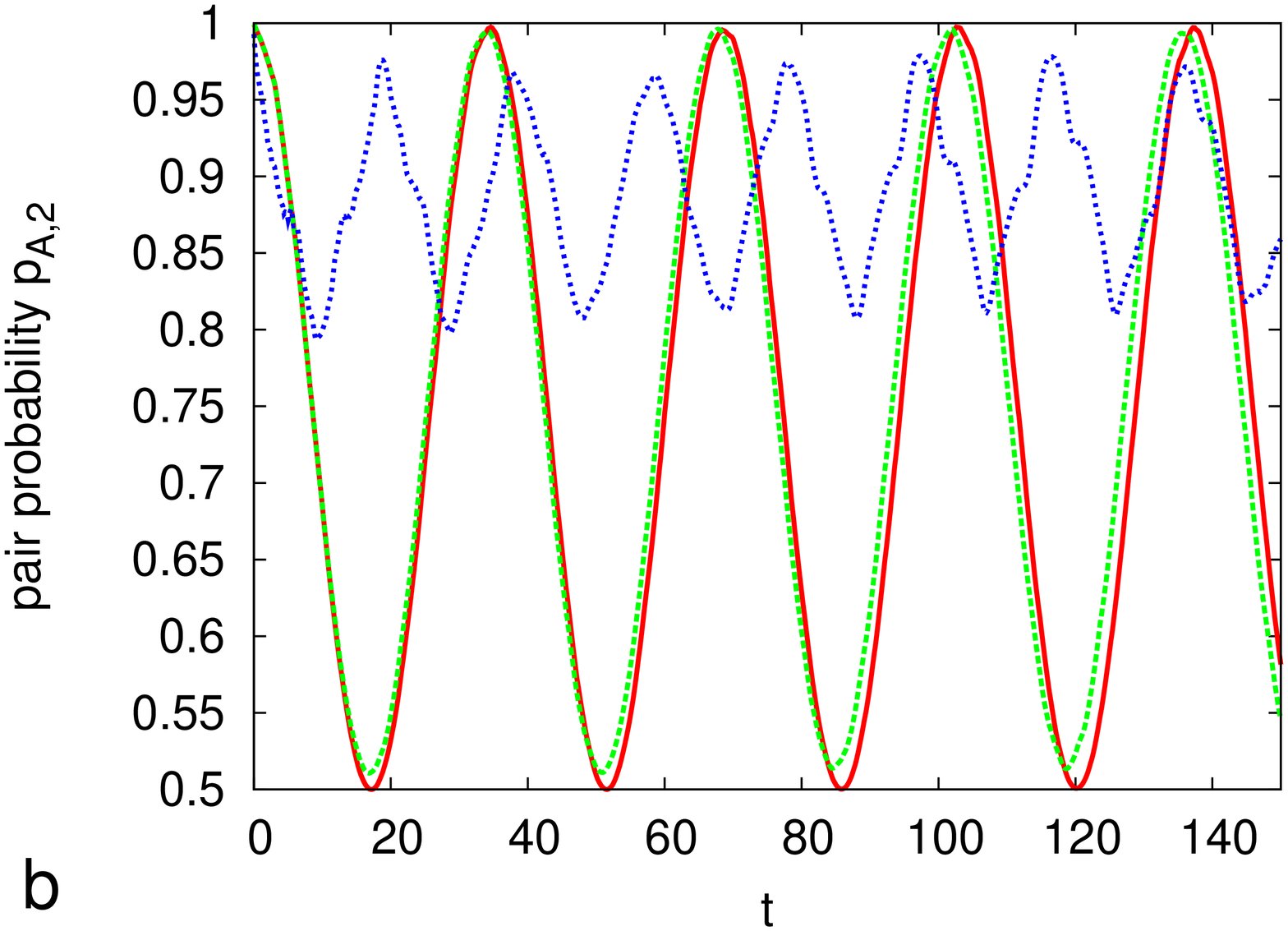}
\caption{(color online) Inter-species tunneling dynamics of noninteracting ($g_{\mathrm{A}}=0$) $N_{\mathrm{A}}=2$ bosons through a strongly repulsive $B$ atom $(g_{\mathrm{AB}}=8)$: \textit{Top:} Evolution of the population of the $A$ bosons for increasing mass ratios $\alpha=0.001, 0.01, 0.122$ \textit{(from left to right)}; \textit{Middle:} Same as above but for the $B$ boson; \textit{Bottom:} (a) Relative population of the right-hand side over time, $p_{\mathrm{A,R}}(t)$; (b) Probability $p_{\mathrm{A},2}(t)$ of finding two $A$ atoms on the same side  for $\alpha=0.001$
(\textbf{\textcolor{red}{---}}), $\alpha=0.01$ (\textbf{\textcolor{green}{-
- -}}), and $\alpha=0.12$ (\textcolor{blue}{${\color{blue}\boldsymbol{\cdots}}$})}
\label{tunnel:ga0}
 \end{figure}

\begin{figure}
\begin{center}
 \includegraphics[width=0.6\columnwidth,keepaspectratio]{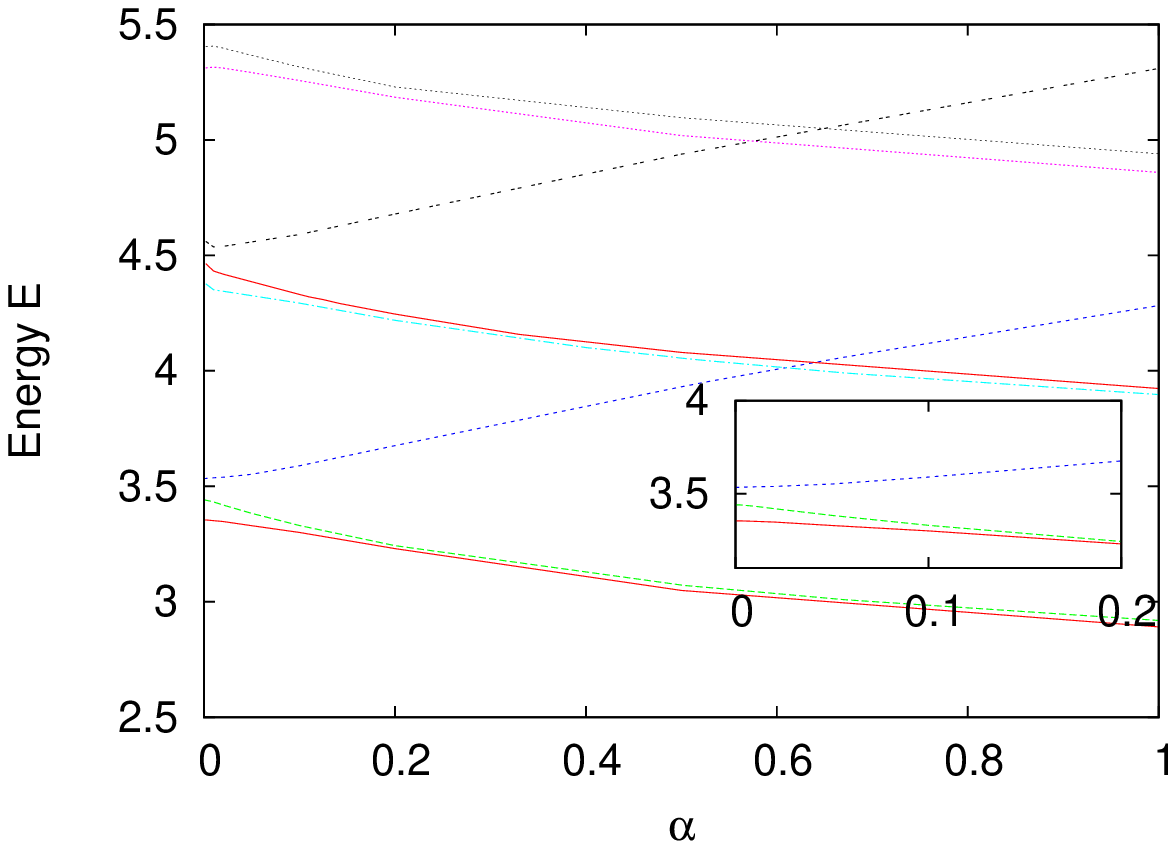}
  \includegraphics[width=0.6\columnwidth,keepaspectratio]{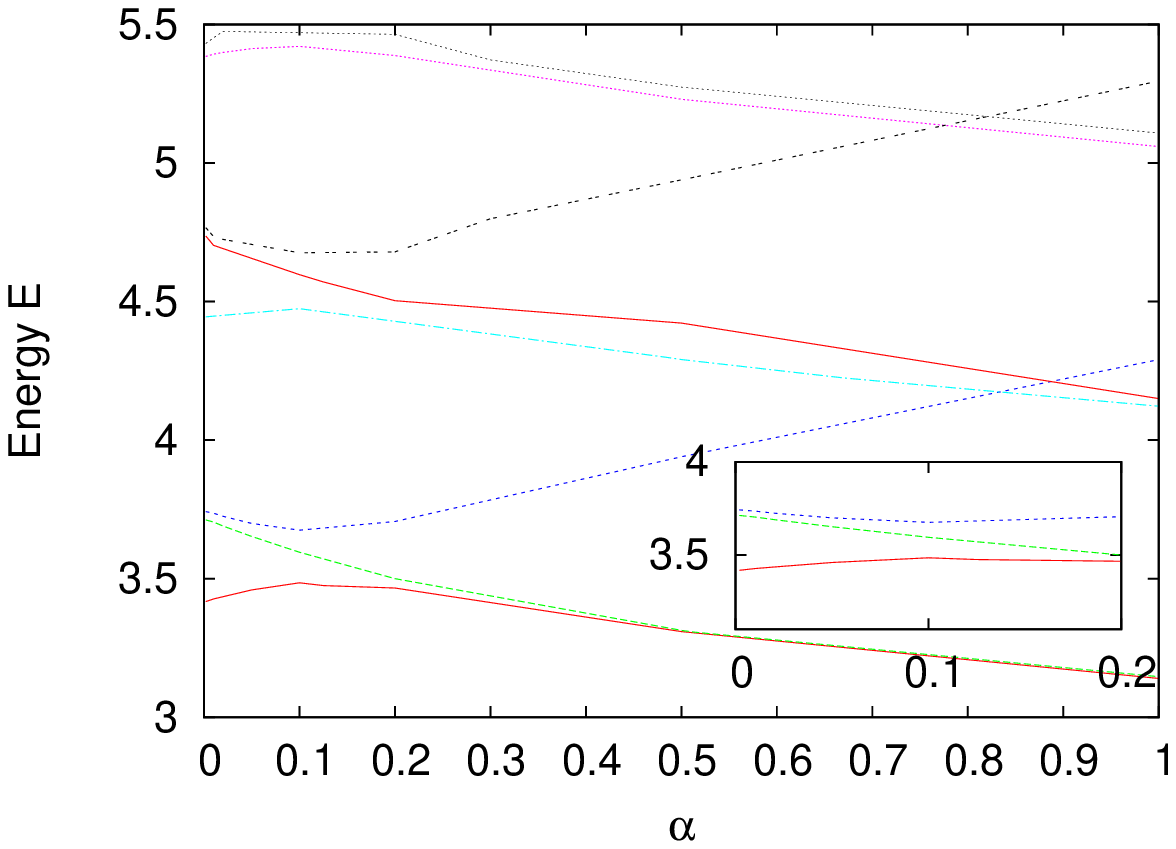}
\includegraphics[width=0.6\columnwidth,keepaspectratio]{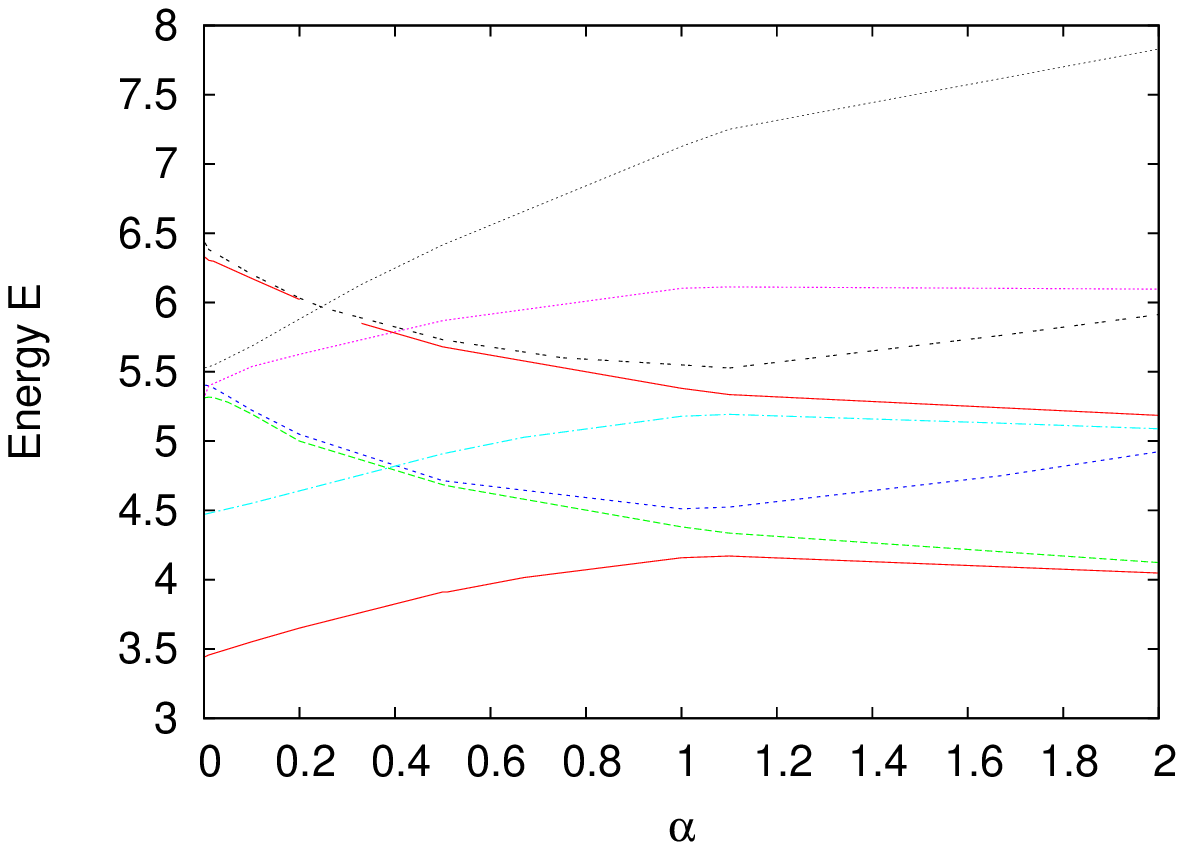}
\caption{(color online) Spectrum for $N_{\mathrm{A}}=2$ and $N_{\mathrm{B}}=1$ bosons with varying $\alpha$; \textit{Top:} $g_{\mathrm{A}}=0$, \textit{Middle:} $g_{\mathrm{A}}=0.5$ \textit{Bottom:} $g_{\mathrm{A}}=25$; $g_{\mathrm{AB}}=25$ throughout}
\label{spectrum_g0}
\end{center}
\end{figure}
We first focus on the case of \textit{noninteracting $A$ bosons} $g_{\mathrm{A}}=0$, as illustrated in Fig.~\ref{tunnel:ga0}. For a vanishing mass ratio $\alpha =0.001$, the $B$ bosons remain fixed at the center of the trap, while we observe simple Rabi-type oscillations between the right and the left side, illustrated by the evolution of the density of $A$ bosons in time (top) and the percentage of atoms at the right side of the trap $p_{\mathrm{A, R}}(t)=\int_0^{\infty} \rho_{\mathrm{A}}(x;t) dx$. 
The observed tunneling time scales become clearer if we consider the spectrum. In the limit of tiny mass ratios $\alpha=0.001$, it is appropriate to use the effective Hamiltonian in eq. \eqref{eq:H_A} as a means to explain the occurring effects. It treats the $B$ bosons as an effective barrier for the lighter $A$ component. The resulting effective potential for the $A$ bosons thus resembles a double well, for which the lowest energy states for $N_{\mathrm{A}}$ bosons in the noninteracting limit, $g_{\mathrm{A}}=0$, are given by distributing all atoms over the symmetric and antisymmetric single-particle orbitals of the lowest doublet (also see \cite{zoellner07b}).  For $N_A$ bosons this leads to $N_A+1$ energies
$E_m=E^B_0 + E^A_0 + 2 m J^{(0)},$
with $m=0,...,N_A$ and $2 J^{(0)}=\epsilon_1^{(0)}-\epsilon_0^{(0)}$ being the energy splitting of the first band. For vanishing interactions and $N_{\mathrm{A}}=2$ particles, only $N_{\mathrm{A}}+1=3$ states contribute and the population of the right-hand side can be easily computed as
\begin{eqnarray}
\nonumber  p_{\mathrm{A, R}}(t)&=&\langle\Psi(t)| \Theta(x_{\mathrm{A}}) | \Psi(t)\rangle\\ \nonumber
&=&\frac{1}{2}( 1 +  c_0 c_1 \langle\Psi_0| \Theta(x_{\mathrm{A}}) | \Psi_1\rangle \cos(\omega_{01}t)\\
&+& c_1 c_2 \langle\Psi_1| \Theta(x_{\mathrm{A}}) | \Psi_2\rangle \cos(\omega_{12}t)),
\label{eq:p_R}
\end{eqnarray}
 where $c_m=\langle\Psi_m| \Psi(0)\rangle$ and the (02) contribution vanishes as only opposite-parity states are coupled. As illustrated in Fig.~\ref{spectrum_g0}, for $g_{\mathrm{A}}=0$ the lowest levels are equidistant and consequently only a single-mode Rabi frequency $\omega_{01}=\omega_{12}=2J^{(0)}$ contributes.
 
Another measure, that gives a better insight into the dynamical properties, is the pair probability
$p_{\mathrm{A},2}(t)=\int_{\left\lbrace x_{1} \cdot x_{2}\geq 0\right\rbrace } \rho_{\mathrm{AA}}(x_{1}, x_{2}; t)\: dx_{1}\:dx_{2},$
which is the integral over the time-dependent two-body correlation function, giving the probability of finding two A bosons at the  same side of the trap.
Noninteracting bosons in a static double well potential tunnel independently as reflected in the two-body correlation function: if both atoms start out in one well, their pair probability will drop from one to one half at equilibrium points of the oscillation as illustrated in Fig.~\ref{tunnel:ga0}.

Increasing the mass ratio, it becomes clear that even for small $\alpha=0.01$ the mobility of the $B$ bosons has dramatic consequences for the tunneling dynamics of the lighter atoms: beats with a frequency $\omega_{12}-\omega_{01}$ begin to evolve on top of the sinusoidal oscillations. It is visible in the spectrum that $\omega_{01}$ is slightly smaller than $\omega_{12}$ and eq. \eqref{eq:p_R} explains the resulting two-mode oscillation. 

For even higher mass ratios $\alpha \approx 0.1$, the tunneling slows down drastically with only a tiny faster modulation on top. 
The form of the spectrum (Fig.~\ref{spectrum_g0}) gives an explanation for this behavior: for $\alpha=0.12$, the lower two lines $E_{0,1}$ are virtually glued together to form a doublet, whereas the gap to $E_2$ increases. This explains the occurring tunneling pattern: complete population transfer occurs at a fairly long time scale, determined by the splitting of the lower states $\omega_{01}$, while $\omega_{12}$ only evokes a small oscillation on top of it.

The effective Bose-Hubbard model \eqref{eq:BHM} explains this phenomenon by introducing an effective attractive intra-species interaction $\delta u<0$ (see Sec.~\ref{effmod}). This attraction makes it more favorable for both bosons to assemble in the same well. 
The initial state, where all bosons are prepared in the left-hand well, is thus only in resonance with a configuration with all bosons in the right-hand well and not resonant with the state where one boson occupies each well. As the bosons cannot gain energy, they cannot tunnel to this energetically higher state, and single-atom tunneling is highly suppressed. In fact, the Bose-Hubbard model suggests the interpretation of the bosons as \textit{attractively bound pairs} \cite{winkler06,foelling07, zoellner07b, zoellner07c}. This picture is also supported by the pair probability $p_{\mathrm{A},2}(t)$ depicted in Fig.~\ref{tunnel:ga0}: the probability of finding two bosons in the same well remains close to one with small modulations, that can be interpreted as attempted single-particle tunneling. Although our effective Bose-Hubbard model (eq. \eqref{eq:BHM}) is not strictly valid in this case anymore, it gives the right tendency for the tunneling times: $T \sim 2 \pi \delta u / 4J^2 \approx 260$, compared to the numerically exact value $T\approx 220$.\\
\textit{The dynamics of the heavier $B$ bosons}\\
Also the dynamics of the $B$ boson becomes more intriguing for an increasing mass ratio, as they begin to move due to the backaction of the lighter tunneling atoms. They remain fixed at the center of the trap in the limit of vanishing mass ratios $\alpha=0.001$ (Fig.~\ref{tunnel:ga0}). For slightly increased $\alpha=0.01$, they carry out tiny oscillations around zero and only begin to move noticeably for larger mass ratios $\alpha \approx 0.1$. Their movement is always in reverse to the $A$ atoms, which is visible comparing their density evolutions. For instance, at $\alpha=0.12$, the lighter bosons tunnel as attractively bound pairs on a long time scale, hence also the dynamics of the $B$ bosons takes place at long scales. This behavior is obvious considering the fact that they are strongly repelled by the $A$ bosons and move in order to avoid them. By extension, this dynamics in reverse to the tunneling atoms is present for all intra-species interactions strengths $g_{\mathrm{A}}$.

\subsection{Intermediate interactions $g_{\mathrm{A}} \sim 0.5$}
\begin{figure}
 \includegraphics[width=0.45\columnwidth,keepaspectratio, angle=270]{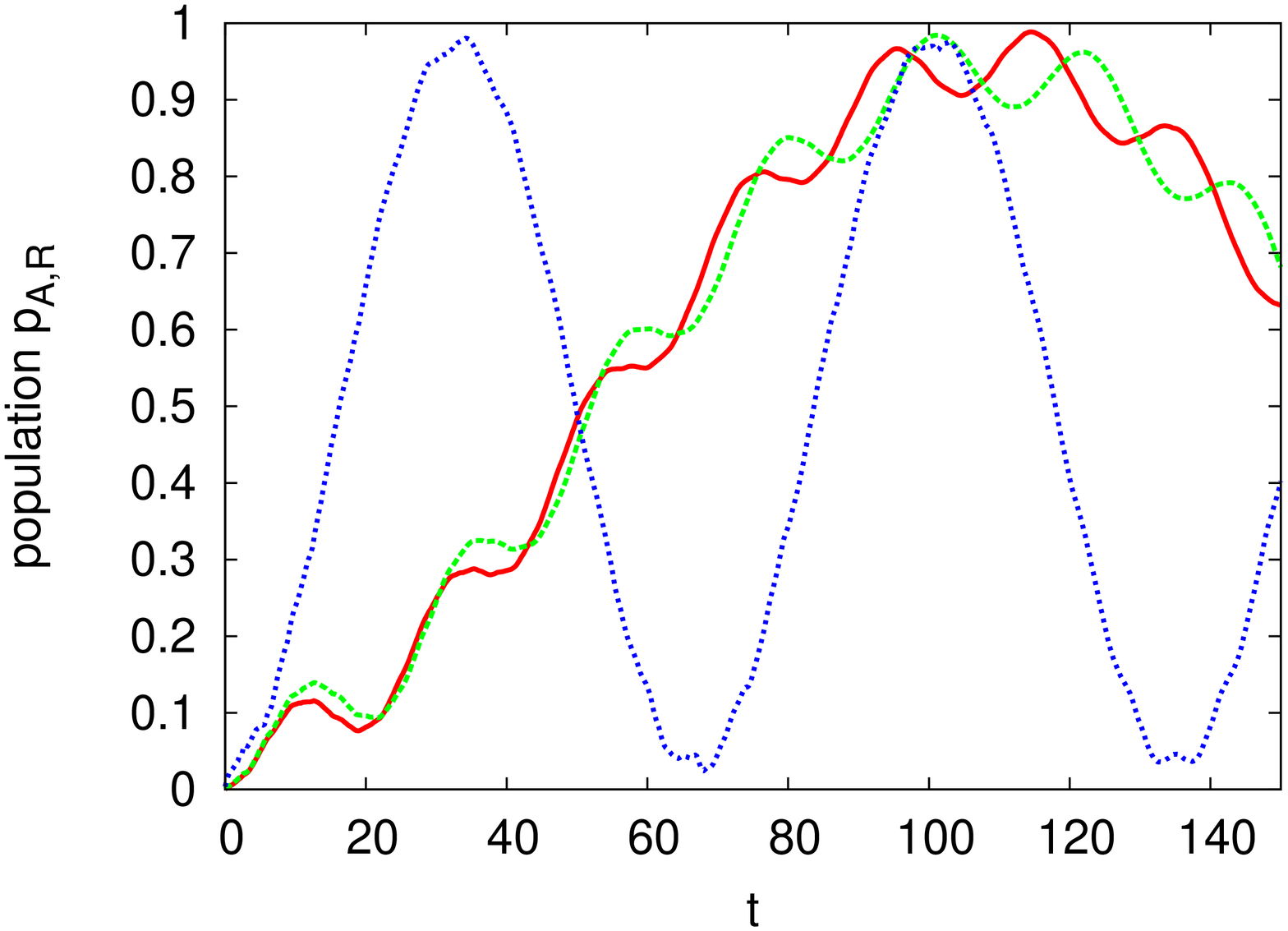}
\includegraphics[width=0.45\columnwidth,keepaspectratio, angle=270]{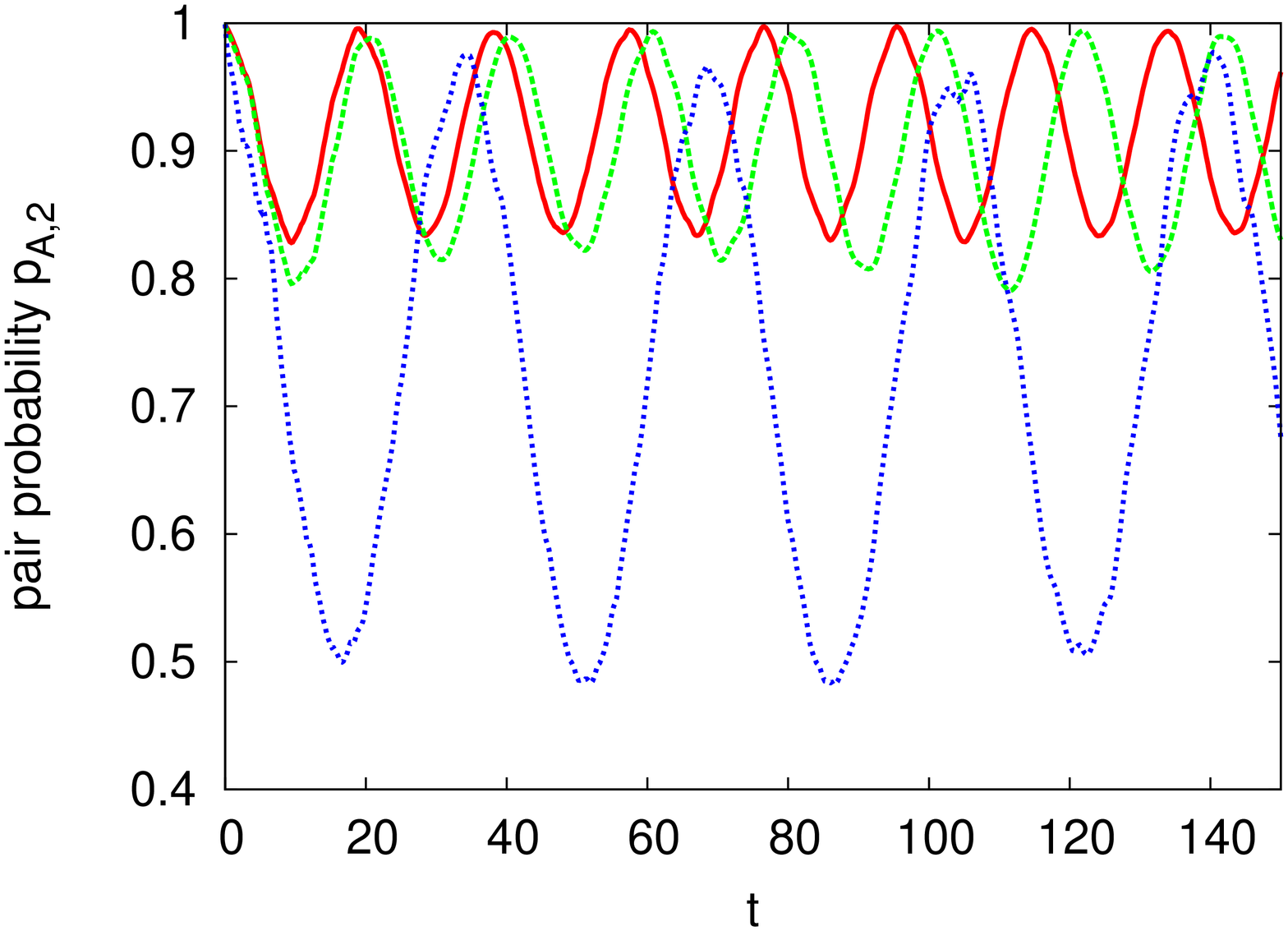}
\caption{(color online) Inter-species tunneling dynamics of $N_{\mathrm{A}}=2$ bosons with intra-species interaction $g_{\mathrm{A}}\sim 0.5$ through a strongly repulsive $B$ atom $(g_{\mathrm{AB}}=8)$ and increasing mass ratios $\alpha=0.001$
(\textbf{\textcolor{red}{---}}), $\alpha=0.01$ (\textbf{\textcolor{green}{-
- -}}), and $\alpha=0.12$ (\textcolor{blue}{${\color{blue}\boldsymbol{\cdots}}$}); \textit{Top:} Relative $A$ population $p_{\mathrm{A,R}}(t)$ and \textit{Bottom}: Pair probability $p_{\mathrm{A},2}(t)$ }
\label{fig:tun_g0.5}
\end{figure}

The case of two \textit{moderately interacting bosons} with $g_{\mathrm{A}}=0.5$ reveals an entirely different behavior as illustrated in Fig.~\ref{fig:tun_g0.5}. Near the static limit ($\alpha=0.001$), the atoms tunnel on very long time scales and only feature a minute faster oscillation on top. This pattern is predicted by the spectrum in Fig.~\ref{spectrum_g0}: two very different time scales determine the dynamical evolution. While the upper two lines $E_{1,2}$ lie very close, leading to a long tunneling time $T=2\pi/\omega_{12}$, the larger splitting of the lower lines $\omega_{01}$ only effects a tiny fast oscillation on top of it. At release time $t=0$, both bosons are prepared in the same well and, due to their additional interaction energy, are off resonance with the configuration, where each well is occupied by one boson. As the system is non-dissipative, no energy can be lost, and it is never possible to access a state with only one boson in each well, although being energetically favorable.
Tunneling of single atoms is therefore strongly suppressed and only the tunneling of pairs is possible.\
The pair correlation $p_{\mathrm{A},2}(t)$ confirms this picture: it roughly remains around 100 $\%$, apart from small reductions due to attempted single-particle tunneling. This indicates that it is always very likely to find both $A$ bosons together either on the left- or the right-hand side of the trap and underlines the stability of the pair. This behavior is reminiscent of the tunneling of repulsively bound atoms pairs, that has been studied for external barriers. \cite{zoellner08b, foelling07}

If we slightly increase the mass ratio $\alpha = 0.01$, the time evolution remains practically the same, as such small changes on the energy levels do not have a visible effect due to the very different tunnel frequencies $\omega_{01}, \omega_{12}$.

For much higher mass ratios $\alpha=0.12$, the repulsively bound pair finally breaks up and the sinusoidal oscillations of the $A$ bosons are recovered. This behavior is predicted by the spectrum in Fig.~\ref{spectrum_g0}, where $\omega_{01} \approx \omega_{12}$ and consequently the dynamics is governed by only one frequency. The effect becomes even more lucid from the angle of the pair probability: instead of remaining around $p_{\mathrm{A},2}(t) \approx 1$, it oscillates between one and one half. These properties are qualitatively predicted within the effective Bose-Hubbard model \eqref{eq:BHM}: the induced on-site attraction $\delta u <0\approx -0.34$ roughly cancels the true intra-species repulsion $u_0>0\approx 0.25$ which leaves us with an effectively noninteracting system that carries out Rabi oscillations with a period $T=\pi/J^{(0)}\approx 65$.
\subsection{The fermionization limit}
\begin{figure}
 \includegraphics[width=0.45\columnwidth,keepaspectratio, angle=270]{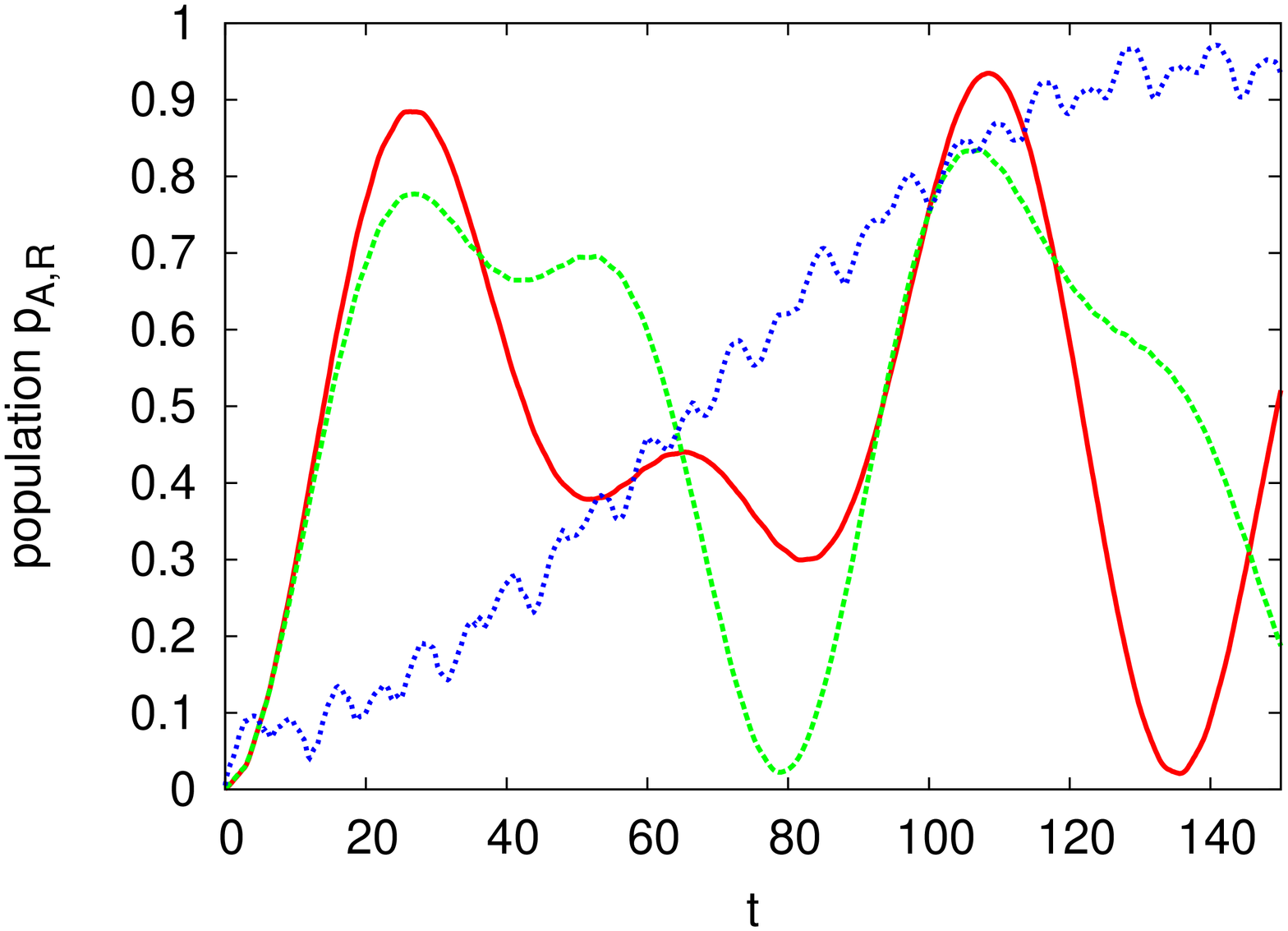}
 \includegraphics[width=0.45\columnwidth,keepaspectratio, angle=270]{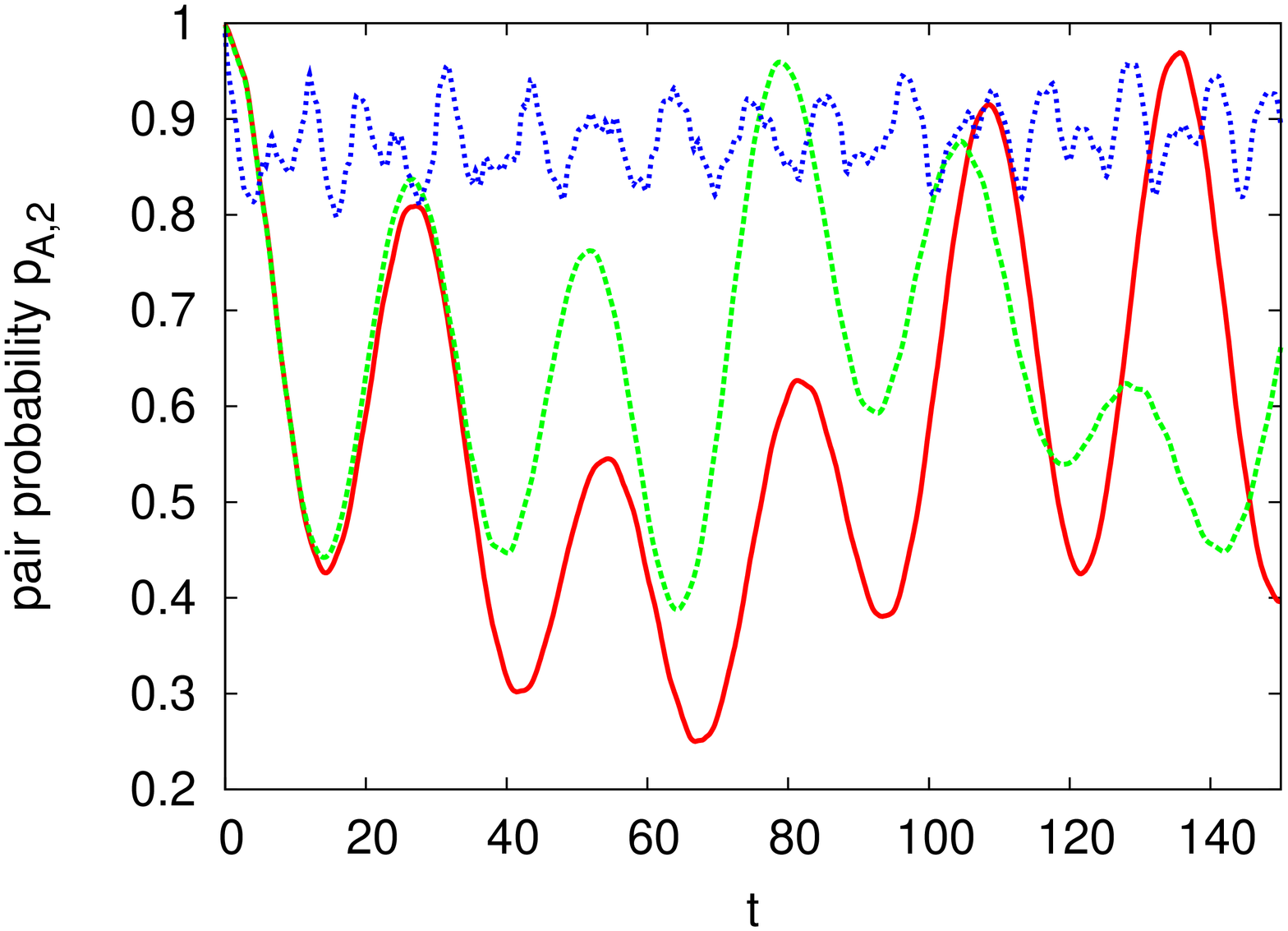}
\caption{(color online) Inter-species tunneling dynamics of $N_{\mathrm{A}}=2$ fermionized bosons ($g_{\mathrm{A}}=25$) through a strongly repulsive $B$ atom $(g_{\mathrm{AB}}=8)$ and increasing mass ratios $\alpha=0.001$
(\textbf{\textcolor{red}{---}}), $\alpha=0.01$ (\textbf{\textcolor{green}{-
- -}}), and $\alpha=0.12$ (\textcolor{blue}{${\color{blue}\boldsymbol{\cdots}}$}); \textit{Left:} Normalized $A$ population $p_{\mathrm{A,R}}(t)$ and \textit{Right:} pair probability $p_{\mathrm{A},2}(t)$ }
\label{fig:tun_g25}
\end{figure}

The dynamics becomes more involved as we increase the intra-species interaction up to the fermionization limit $g_{\mathrm{A}}=25$. Needless to say that in this strongly correlated limit the lowest band approximation as stated in eq.~\eqref{eq:p_R} breaks down and we have to find another way to explain the dynamics. According to the Bose-Fermi map \cite{girardeau60}, hard-core bosons and fermions have an identical energy spectrum and we seek to understand the tunneling dynamics of the bosons by effectively treating them as fermions. As a consequence of the Pauli principle, (spinless) fermions in one well cannot occupy the same band, and consequently populate the $N_{\mathrm{A}}$ lowest bands in the initial state. The probability of finding a particle at the right-hand side can therefore be written as
\begin{equation}
 p_{\mathrm{A,R}}(t)=\frac{1}{N_A}\sum_{\beta=0}^{N_A-1} \sin^2(J^{(\beta)}t) .
\label{fermtun}
\end{equation}
As visualized in Fig.~\ref{fig:tun_g25} for $N_{\mathrm{A}}=2$, the only two frequencies that contribute for $\alpha \approx 0$ are $J^{(0)}$ and $J^{(1)}$. Also the pair probability $p_{\mathrm{A},2}(t)$ supports the analogy between hard-core bosons and fermions: it takes nearly all values between one and zero, pointing out that the dynamics can be understood as the tunneling of two independent fermions with different frequencies.

Recalling the effective Fermi-Hubbard model (eq.~\eqref{FHM}), with increasing $\alpha$ we expect an induced attraction between the tunneling fermionized bosons. In the intermediate
regime ($\alpha=0.01$), this has only a small effect leading to
a renormalization of the contributing Rabi frequencies $J^{(\beta)}$ with $\beta=0,1$ for $N_{\mathrm{A}}=2$.

However, for a strongly increased mass ratios $\alpha=0.12$, remarkable effects emerge: single-particle tunneling is turned off-resonant and the two atoms move only on a time scale about four
times longer than the Rabi oscillations. The tunnel dynamics closely resembles the one of a bound pair, a point of view that is supported by the pair probability $p_{\mathrm{A}, 2}(t)$ that stays remarkably close to unity, in marked contrast with the independent fermion tunneling for $\alpha \approx 0$. In this light, it is tempting to think of the phenomenon as the tunneling of an \textit{attractively bound pair of identical fermions}. The effective Fermi-Hubbard model \eqref{FHM} underpins this interpretation: the induced inter-band interaction $\delta v^{(0101)}$ is always negative, signifying that the fermionized bosons attract each other and only pair tunneling is resonant. While the effective model predicts a tunneling period of $T= 2\pi \delta v^{(0101)}/4J^{(0)}J^{(1)} \approx 320$, the numerically exact result is $T\approx 290$.

\subsection{Higher atom numbers}\label{highernumb}
Although having concentrated on the special case of $N_{\mathrm{A}}=2$ light bosons so far, it would be interesting to consider higher atom numbers for several reasons. On the one hand, in a setup consisting of a whole array of 1D traps like described in \cite{paredes04, kinoshita04, kinoshita06}, number fluctuations may automatically admix states with $N_{\mathrm{A}}>2$. On the other hand we want to verify the validity of the effective Hubbard models for higher atom numbers. 
Furthermore, it would be interesting to connect the dynamics of our few-boson system to that of large atomic clouds. This shall first be discussed for an increased number of $B$ bosons and thereafter for more $A$ atoms.\\

\begin{figure}
\includegraphics[width=0.7\columnwidth,keepaspectratio]{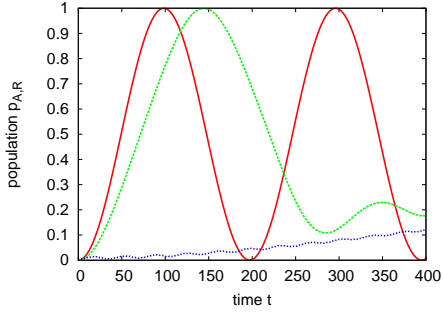}
\caption{(color online) Inter-species tunneling dynamics of $N_{\mathrm{A}}=2$ bosons through $N_{\mathrm{B}}=2$ atoms for $(g_{\mathrm{AB}}=8), g_{\sigma}=0$ and increasing mass ratios: Normailzed $A$ population $p_{\mathrm{A, R}}(t)$ for
$\alpha=0.001$ (\textbf{\textcolor{red}{---}}), $\alpha=0.02$ (\textbf{\textcolor{green}{-
- -}}), and $\alpha=0.12$ (\textcolor{blue}{${\color{blue}\boldsymbol{\cdots}}$})}
\label{more_B}
\end{figure}

\begin{figure}
\includegraphics[width=0.6\columnwidth,keepaspectratio]{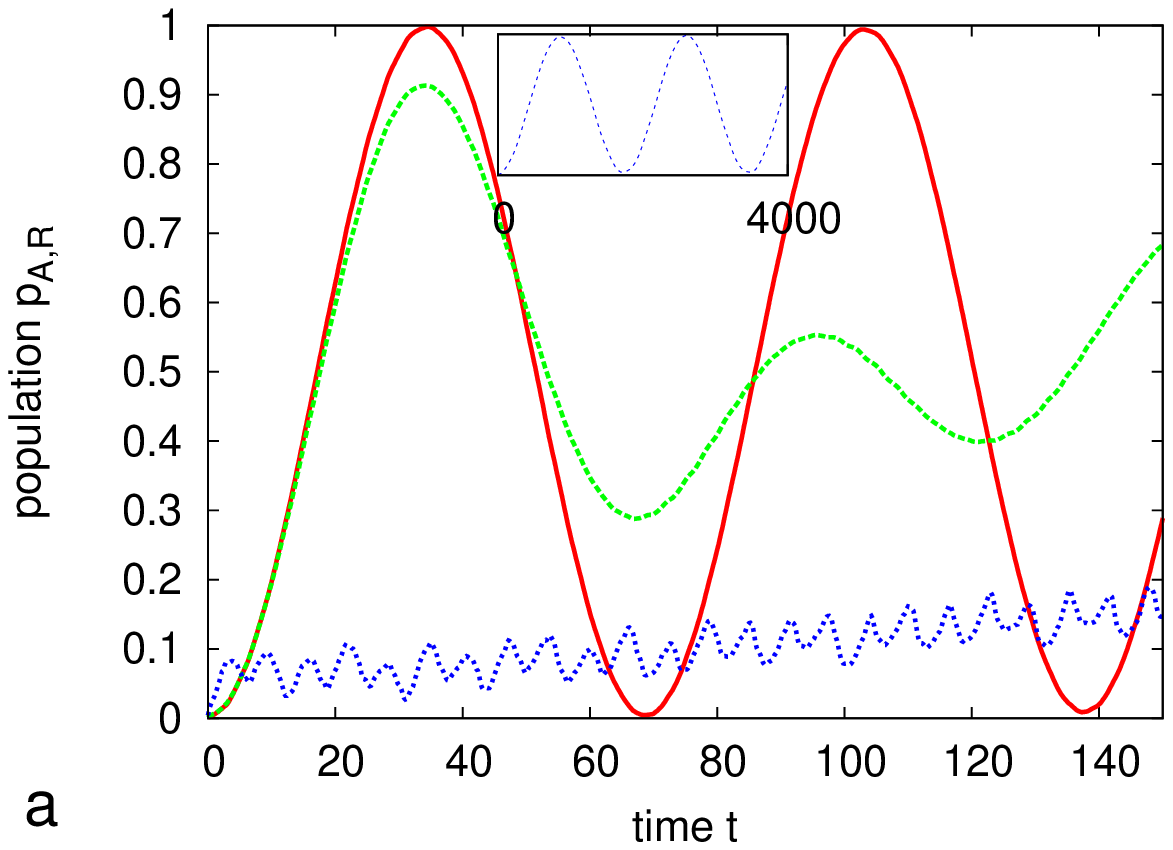}
\includegraphics[width=0.6\columnwidth,keepaspectratio]{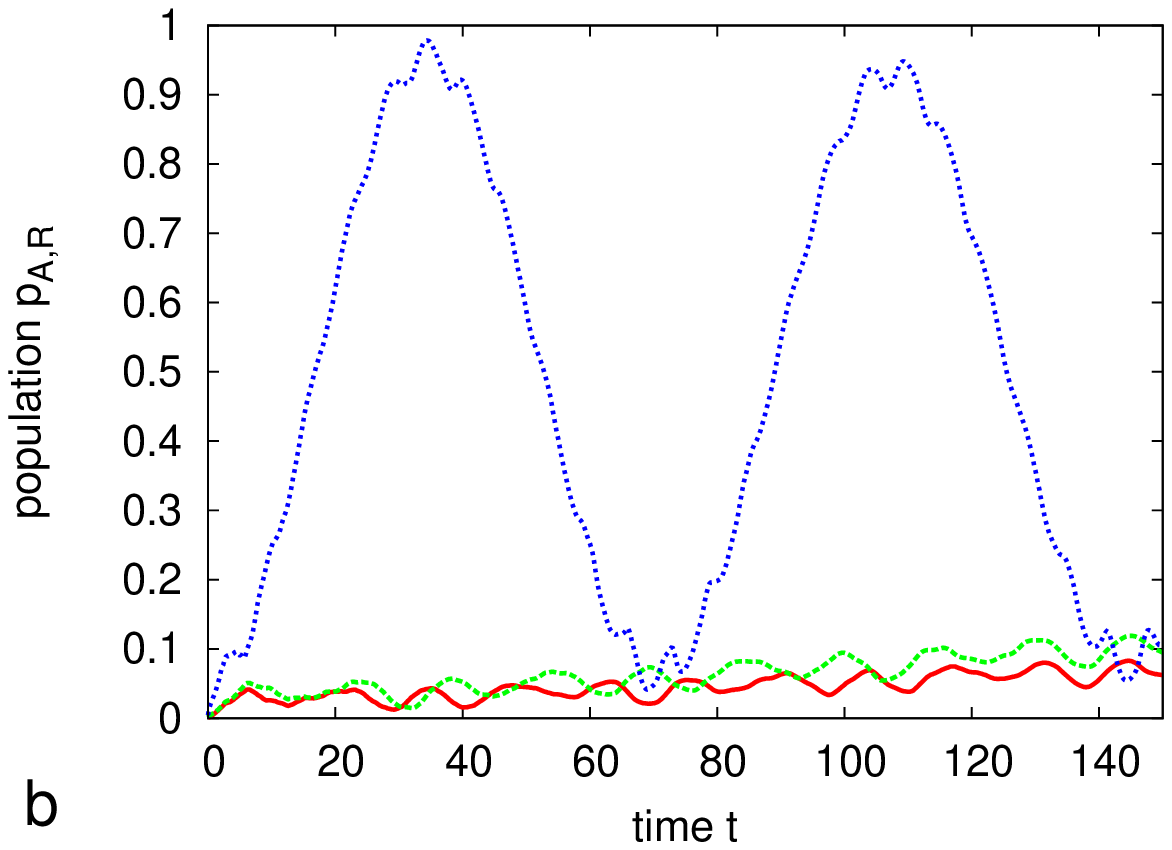}
\includegraphics[width=0.6\columnwidth,keepaspectratio]{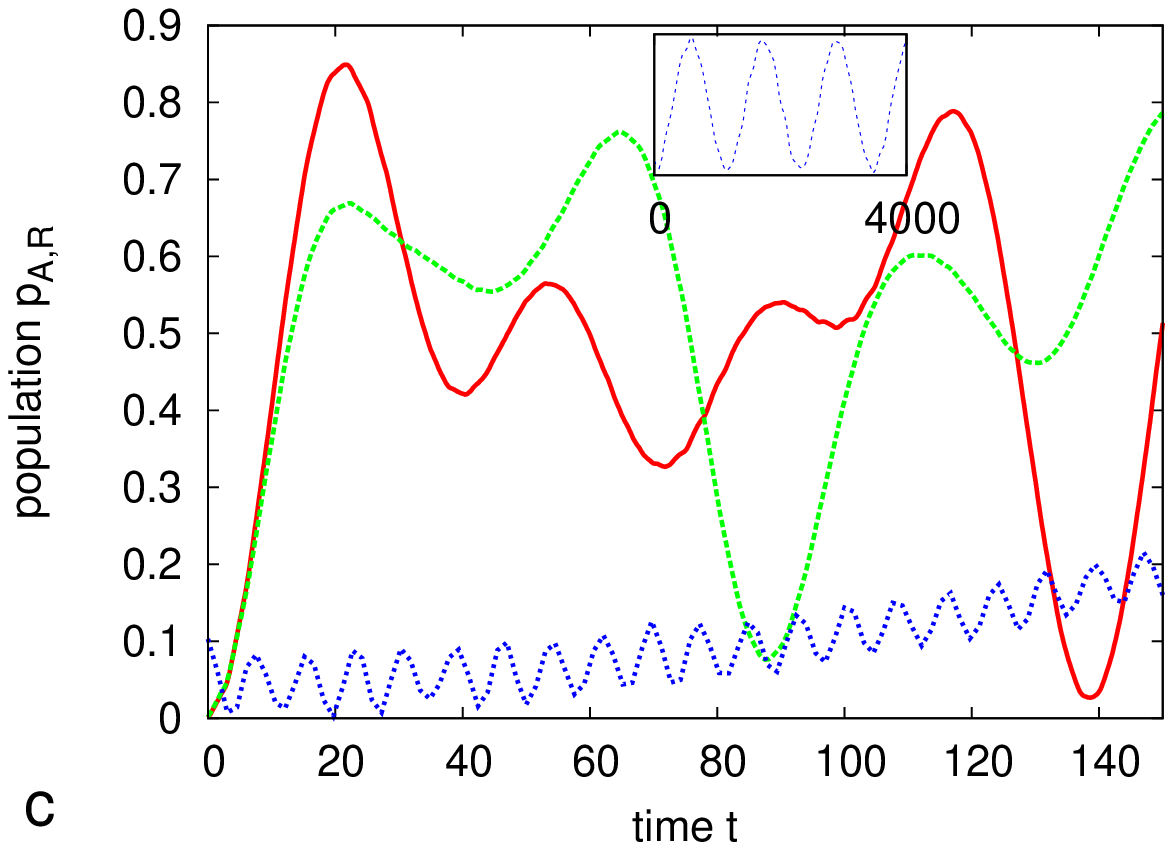}
\caption{(color online) Inter-species tunneling dynamics of $N_{\mathrm{A}}=3$ bosons through a strongly repulsive $B$ atom $(g_{\mathrm{AB}}=8)$ and increasing mass ratios: Relative $A$ population $p_{\mathrm{A,R}}(t)$ for
$\alpha=0.001$ (\textbf{\textcolor{red}{---}}), $\alpha=0.01$ (\textbf{\textcolor{green}{-
- -}}), and $\alpha=0.12$ (\textcolor{blue}{${\color{blue}\boldsymbol{\cdots}}$}), (a): $g_{\mathrm{A}}=0$, \textit{Inset}: Long-time behavior $\alpha=0.12$, (b): $g_{\mathrm{A}}=0.5$, (c): $g_{\mathrm{A}}=25$}. 
\label{Fig:tun_N3}
\end{figure}

\textit{Higher numbers of $B$ bosons:}\\
Increasing the number of $B$ bosons does not change the tunneling dynamics conceptually. In general, the tunneling times are enhanced and the $B$ bosons are more inert in their reaction to the tunneling $A$ bosons due to their increased total mass. In the static limit, $\alpha=0.001$, the heavier atoms display an effective barrier for the lighter species, which is proportional to the number of $B$ bosons, rendering the tunnel coupling $J$ between different sites weaker. The tunnel period $T= \pi/J^{(0)}$ is accordingly enhanced (see Fig.~\ref{more_B}). For an increased mass ratio $\alpha=0.01$ (not shown), the beatings known from $N_{\mathrm{B}}=1$ are not observable: the total mass of the $B$ bosons is larger, which results in a weaker backaction to the tunneling component. Hence the beatings can only be recovered for larger $\alpha=0.02$ as displayed in Fig.~\ref{more_B}. In the limit of larger mass ratios $\alpha=0.12$, we again observe tunneling of attractively bound pairs of $A$ bosons on a longer time scale $T \sim 2 \pi\delta u/4J^{(0)2}$ due to the smaller tunnel coupling $J^{(0)}$.\\
\textit{Higher numbers of $A$ bosons:}\\
The \textit{weakly-interacting} behavior of an increased number of $A$ bosons does not differ conceptually from the case of lower particle numbers. In the limit of $\alpha\rightarrow0$, where the $B$ bosons can be treated as static barrier for the $A$ bosons, eq.~\eqref{eq:p_R} carries over to
\begin{equation*}
  p_{\mathrm{A,R}}(t)= \frac{1}{2} \left( 1+\sum_{m<n}^{N_\mathrm{A}}c_m c_n \langle \Psi_m| \Theta(x) | \Psi_n \rangle \cos(\omega_{nm}t)\right) ,
\end{equation*}
where $N_{\mathrm{A}}$ modes have to be taken into account. For vanishing intra-species interactions $g_{\mathrm{A}}=0$ and a vanishing mass ratio $\alpha \approx 0$, the same simple Rabi oscillations as for smaller atom numbers arise (see Fig.~\ref{Fig:tun_N3}). As we slightly increase the mass ratio to $\alpha=0.01$, the equidistance between the contributing levels is again lifted and beats evolve on top of the sinusoidal oscillations.

As we move to larger mass ratios $\alpha=0.12$, the lowest-lying of the $N_{\mathrm{A}}$ states have merged into doublets, where the highest eigenstate pair is roughly described by $\ket{\Psi_{\mathrm{A}}}=\frac{1}{\sqrt{2}}( \ket{N_L, 0_R} \pm \ket{0_R, N_L})$. This effect results from the induced intra-species attraction $\delta u$ and can be extrapolated from the behavior for $N_{\mathrm{A}}=2$ in Fig.~\ref{spectrum_g0}. Their energy splitting in this two-mode model has been determined to be $\omega \propto \frac{2N_A  u}{(N_A-1)!}\left( \frac{2 J^{(0)}}{u}\right) ^{N_A}$ \cite{salgueiro06}.
The tunnel period is thus expected to grow exponentially with the number of particles as already foreshadowed in Fig.~\ref{Fig:tun_N3} for the case of $N_A=3$ bosons. Extrapolating this behavior to higher particle numbers, we finally reach the condensate regime ($N_A \gg 1$), where complete population transfer becomes completely inaccessible.
The same prediction is made within the effective Bose-Hubbard model that yields a tunneling period of $T\approx 2 \pi u^2/3J^{(0)3} \approx 2500$ for three $A$ bosons for a mass ratio $\alpha=0.12$, compared to the numerically exact value $T \approx 1800$, depicted in the inset in Fig.~\ref{Fig:tun_N3}(a).\\

The same scaling of the tunneling period can be observed for an intrinsically interacting system in the case of intermediate coupling strengths $g_{\mathrm{A}}\approx 0.5$: for $N_{\mathrm{A}}=3$, we observe long tunneling times that increase exponentially with the number of $A$ bosons. Only tiny corrections are remarkable for small increments of the mass ratio ($\alpha\approx 0.01$). It is only for larger mass ratios $\alpha \approx 0.12$ that the induced attraction $\delta u$ is sufficiently strong to balance the true repulsion $u_0$, and we recover Rabi oscillations, already known from the case of $N_{\mathrm{A}}=2$ bosons in Fig.~\ref{Fig:tun_N3}(b). This behavior can be extrapolated to larger particle numbers, where the induced attraction $\delta u$ always cancels $u_0$ at approximately $\alpha \approx 0.12$.\\

For \textit{strong intra-species repulsions $g_{\mathrm{A}}=25$}, the fermionized bosons can again be mapped onto noninteracting fermions and eq.~\eqref{fermtun} yields the proper description. For $N_{\mathrm{A}}=3$ hard-core bosons, three Rabi frequencies contribute, explaining the picture that emerges in the limit of vanishing mass ratios $\alpha \approx 0$ (see Fig.~\ref{Fig:tun_N3}(c)). In the intermediate regime ($\alpha=0.01$), beatings on each of the contributing "Rabi frequencies" $J^{(\beta)}$ evolve, which do not qualitatively alter the overall tunneling dynamics. For higher mass ratios $\alpha=0.12$, we can again observe the tunneling of attractively bound pairs of fermions, which is properly accounted for in the effective Fermi-Hubbard model. Due to energy conservation, only terms that conserve the number of atoms in each band contribute to the inter-species tunneling dynamics, such that the on-site interaction Hamiltonian reduces to $H_{\mathrm{on-site}}=\sum_{s, \alpha < \beta} \delta v^{(\alpha \beta\alpha\beta)} \hat{n}_s^{\alpha}\hat{n}_s^{\beta}$. All contributions for $N_{\mathrm{A}}=3$, namely $\delta v^{(0101)}, \delta v^{(0202)}, \delta v^{(1212)}$,  are negative and lead to a tunneling time  of $T\approx 1600$, compared to the numerically exact result $T\approx 1150$.

\section{Conclusion}
In summary, we have studied two-component mixtures of one-dimensional few-boson systems under harmonic confinement. In our setup, the two species experience very different confinement lengths, realized via different masses (in analogy, this can also be achieved by a variation of the trapping frequencies). By varying the mass ratio, different regimes, from the complete localization of one species at the center of the trap to equal masses for both constituents, have been investigated. The inter-species interaction has been fixed at a large repulsive value, while the full crossover from weak intra-species coupling to the fermionization limit has been investigated for the lighter bosons.\\
 In the limit where the two species experience a very different degree of localization, the heavier one can be considered as an effective barrier for the lighter species and we can map this to a system of identical bosons in a double well. On easing the localization requirement of the heavier bosons, an additional correlation between the lighter species arises. For this case, we rigorously derive an effective lattice model and show that one can account for the mobility of the barrier in terms of an induced \textit{attraction}.  For the ground state properties, this may lead to strong intra-species correlations even in the case of intrinsically noninteracting bosons, resulting in an increased probability of finding several atoms at the same position. On the other side, it is possible to cancel an initial intra-species repulsion by means of the induced attraction, leading us back to an effectively uncorrelated system. Also the tunneling dynamics of the mobile component through the effective barrier reveals a similar physical behavior. In the limit of complete localization of the heavier species, the dynamics through the effective barrier bears a strong resemblance to tunneling in a double well. Increasing the mass ratio, it is possible to go from Rabi oscillations in a system with vanishing intra-species interactions, to a regime, where only tunneling of attractively bound pairs is possible. Notably, this is even realizable for fermionized bosons, that, despite of being strongly repulsive, begin to tunnel in attractively bound pairs through the heavier species.
This dynamical behavior is not limited to systems with only one impurity and by extension, it would also be possible to create and investigate an effective lattice of localized atoms, allowing for the study of highly-correlated quantum systems.

\appendix
\section{Computational Method}\label{mctdh}
The results described in this paper are based on the numerically exact Multi-Configuration Time-Dependent Hartree method \cite{meyer90, meyer99} that has been successfully applied to few boson systems \cite{zoellner06a, zoellner06b, zoellner07a, zoellner07b, zoellner07c, tempfli08, tempfli09, eckhart09}. Its principal idea is to solve the many-body Schr{\"o}dinger equation $i \dot{\Psi}(t)= H\Psi(t)$ as an initial-value problem by expanding the solution of the wave function in terms of direct (Hartree) product states $\Phi_J(t)=\varphi_{j_1}(t)\otimes ..\otimes \varphi_{j_N}(t) $
\begin{equation}
 \Psi(t)=\sum_{J \in C}A_J(t) \Phi_J(t),
\label{eqmot}
\end{equation}
where the multiindex $J=(j_1, ..., j_N)$ runs over all admissible configurations $C=\left\lbrace J=(j_1, ..., j_N)|1 \leq j_i \leq s\right\rbrace$ with $s$ denoting the maximum number of required basis functions. The single-particle functions $\left\lbrace \varphi_j(t)|1\leq j\leq s\right\rbrace$ that are used to built up the product states are still unknown and represented in a fixed \textit{primitive} basis on a grid, which is in our case provided by harmonic oscillator functions.

Applying the Dirac-Frenkel variational principle yields equations of motion for both the coefficients $A_J(t)$ and the single-particle functions $\varphi_j(t)$. Integrating this system of differential equations leads to the time evolution of the system via \eqref{eqmot}. In the above expansion not only the coefficients $A_J(t)$, but also the single-particle functions $\varphi_j(t)$ are time dependent, which results in a basis set $\left\lbrace \Phi_J(t) \right\rbrace$ that is variationally optimal at each time $t$ and can thus be kept relatively small, lending this method its high efficiency. Still, the exponential growth of the numerical effort limits the approach to particle numbers $N<10$ depending on the number of single-particle functions necessary to describe the inter-particle correlations.\ 
The peculiarity of the system at hand, a two-component mixture of bosons, is the indistinguishability within each species, which manifests in identical single-particle functions within each subset $ K_A= \left\lbrace 1,..., N_A\right\rbrace $ and $K_B=\left\lbrace  N_A+1,...,N\right\rbrace$. The bosonic permutation symmetry is not a priori included in the ansatz \eqref{eqmot} and is ensured by the correct symmetrization of the expansion coefficients $A_J$. 
 
Besides time-dependent dynamical investigations, this paper also contains stationary-state calculations, which can be obtained employing the so-called \textit{relaxation} method \cite{kosloff1986}. The principal idea behind this approach is to propagate an initial wave function $\Psi(0)$ by the non-unitary, imaginary time-propagator $U(\tau)=e^{-H\tau}$. For $\tau \rightarrow \infty$, this damps out any contribution from excited states by the exponential factor $e^{-(E_M-E_0)\tau}$ and we are left with the ground state. This basic principle is very sensitive to numerical instabilities, and the scheme that one relies on in practice is the more robust \textit{improved relaxation} \cite{worth2003}. The starting point of this ansatz is to minimize $\langle \Psi | H | \Psi \rangle$ with respect to both the coefficients $A_J$ and the orbitals $\varphi_j$. The resulting eigenvalue problem is solved iteratively by first diagonalizing the Hamiltonian in some \textit{fixed} basis to obtain the coefficients $A_J$. The orbitals $\varphi_j$ are then optimized by propagation over a short period in imaginary time and the Hamiltonian matrix is subsequently rebuilt with these new orbitals. This procedure is then repeated until the desired accuracy is obtained. 

\begin{acknowledgments}
Financial support from Landesstiftung Baden-W\"urttemberg through the project \textit{Mesoscopics and atom optics of small ensembles of ultracold atoms}  and from the German Academy of Science Leopoldina (grant LPDS 2009-11; S.Z.) is gratefully acknowledged. The authors thank H.-D.~Meyer,
S.~Jochim, T.~Gasenzer, A.~Jackson and T.~Barthel for valuable discussions.

\end{acknowledgments}

\bibliographystyle{prsty}
\bibliography{main}

\end{document}